\documentclass[12pt, onecolumn]{article}
\usepackage{amsmath, amsthm, amssymb}
\usepackage[margin=1in]{geometry}
\usepackage{float}
\usepackage{constants, color, mathtools, tikz, xcolor}
\usepackage{amssymb,amsmath,amsthm,graphicx,hyperref}
\usepackage{inputenc} 
\usepackage{rotating}
\usepackage{pgfgantt} 
\usetikzlibrary{shapes,snakes}
\usepackage{authblk}
\usepackage[linesnumbered,ruled,vlined]{algorithm2e}
\usepackage{setspace}
\usepackage{booktabs} 
\usepackage{multirow} 
\usepackage{array} 
\usepackage{makecell} 
\usepackage{subcaption} 
\usepackage{soul} 
\usepackage{url}
\usepackage{thmtools} 
\usepackage{thm-restate} 
\usepackage[labelfont=bf]{caption} 
\usepackage{natbib} 
 \bibpunct[, ]{(}{)}{,}{a}{}{,}

\SetKwInput{KwInput}{Input}                
\SetKwInput{KwOutput}{Output}              
\SetKw{KwBreak}{break}
\SetKw{KwContinue}{continue}
\SetKw{KwNot}{not}
\SetKw{KwAnd}{and}
\SetKw{KwOr}{or}
\SetKw{KwTrue}{true}
\SetKw{KwFalse}{false}

\graphicspath{ {./images/} } 
\usepackage{float} 

\newcommand{\erdosrenyi}{Erd\H{o}s--R\'enyi}

\newtheorem{thm}{Theorem}[section]

\begin{document}

\let\WriteBookmarks\relax
\def\floatpagepagefraction{1}
\def\textpagefraction{.001}

\numberwithin{figure}{section}

\title{Perfect Graph Modification Problems: An Integer Programming Approach}



\author{Burak Nur Erdem\footnote{E-mail: \texttt{burak-nur.erdem@lirmm.fr}}}
\author{T{\i}naz Ekim\footnote{E-mail: \texttt{tinaz.ekim@bogazici.edu.tr}}}
\author{Z. Caner Taşkın\footnote{E-mail: \texttt{caner.taskin@bogazici.edu.tr}}}
\affil{\normalsize Department of Industrial Engineering, Bo\u{g}azi\c{c}i University, 34342, Bebek, Istanbul, Turkey.}


\date{\vspace{-6ex}}
\maketitle

\begin{abstract}
Graph modification problems aim to find a small set of modifications to a graph so that it satisfies a desired property. The literature is rather rich in NP-completeness results and polynomial time solvable cases for special graph classes. However, no exact algorithm has been proposed for perfect graph modification problems. In this work, we propose the first exact solution methods based on integer programming for three variants: minimum perfect editing, minimum perfect completion, and the perfect sandwich problems. The minimum perfect editing problem inquires about the smallest number of edge additions and deletions needed to make a graph perfect, while the completion problem allows only for edge additions. The perfect sandwich problem is a decision problem that asks whether a perfect graph can be formed by adding edges from a restricted subset. To solve these problems, we formulate an integer programming model based on the Strong Perfect Graph Theorem. To address the resulting exponential number of constraints, we propose a branch-and-cut algorithm that dynamically generates them on demand. At the core of this approach is an efficient separation routine for enumerating odd holes and odd antiholes. We also release this underlying routine as \texttt{is\_perfect}—a standalone open-source perfect graph recognizer and odd hole enumerator designed for broader community reuse. To enhance the practical efficiency of the branch-and-cut algorithm, we calculate the expected number of odd holes and odd antiholes in random \erdosrenyi{} graphs. In addition, we propose \texttt{IterativeModificationHeuristic}, the first heuristic for the editing and completion problems, which provides upper bounds. Finally, we demonstrate the empirical effectiveness of the proposed methods through computational experiments on a wide range of instance types; all benchmark instances are publicly available.
\end{abstract}




\section{Introduction}



Graph modification problems aim to transform a given graph to satisfy a target property through a minimum number of edge additions (completion), deletions, or both (editing). A related variant, the graph sandwich problem, asks whether a graph satisfying a property can be obtained by adding edges exclusively from a predefined set of optional non-edges; if this optional set is empty, it reduces to the standard recognition problem.

Perfect graphs arise naturally in applications such as scheduling and frequency assignment \citep{Golumbic}. In practice, input graphs often fail to satisfy desired structural properties due to noise or modeling constraints. Perfect graph modification provides a quantitative measure of this structural distance. For instance, in wireless frequency assignment, modeling conflicts as a vertex coloring problem on a perfect graph ensures tractability \citep{Golumbic}. If the graph is not perfect, removing a minimum number of edges to achieve perfection identifies a small set of tolerated interferences, explicitly balancing model fidelity with computational efficiency.

Graph modification problems have wide-ranging applications across circuit design \citep{elmallah_colbourn}, numerical linear algebra \citep{ROSE1972183, yannakakis_chordal}, molecular biology \citep{cirino1997, goldberg_golumbic_four_strikes, golumbic_kaplan_shamir_dna_mapping}, chemistry \citep{bip_edit_dist_molecular_chem}, and machine learning \citep{bansal_blum_chawla}. Although most of the literature focuses on classical complexity \citep{np_completeness_modification, sandwich_almost_monotone, np_completeness_sandwich, sandwich_1join_npc, survey_complexity_modification, elmallah_colbourn, natanzon_shamir_sharan, np_completeness_modification2} and parameterized complexity \citep{chordal_editing_is_fpt, survey_complexity_modification, parameterized_complexity_on_degree_edit, chordal_sandwich, chordal_deletion_is_fpt}, exact integer programming methods exist for problems such as graph edit distance \citep{bip_edit_dist_molecular_chem, bip_graph_edit_distance}, cluster editing \citep{ip_cluster_edit, ip_cluster_edit2}, line-invertible deletion \citep{IP_line_deletion}, and chordal graph completion \citep{chordalcompletion}. Our approach aligns most closely with a related study on chordal completion \citep{chordalcompletion}, which utilizes an IP formulation and a branch-and-cut algorithm. However, due to differing target graph classes, their models are not directly transferable and require modifications that exploit perfect graph properties. Similarly, \citet{InducedBip} proposes a branch-and-cut algorithm detecting odd cycles for the bipartite induced subgraph problem. The fundamental nature of the two problems differs; as their approach relies on vertex set operations, ours focuses on modifying the edge and non-edge sets.

The minimum perfect editing, completion, and deletion problems are NP-hard \citep{natanzon_shamir_sharan}. The complexity of the perfect sandwich problem itself remains an open question, although closely related odd-hole-free and even-hole-free sandwich problems were recently shown to be NP-complete \citep{sandwich_odd_hole_even_hole}.

To the best of our knowledge, no exact or heuristic algorithms currently exist for perfect graph modification problems. This paper bridges this gap by proposing the first exact branch-and-cut algorithm for the minimum perfect editing, completion, and sandwich problems. Additionally, we introduce a heuristic algorithm that provides strong upper bounds for the exact method and serves as an efficient standalone approach. The proposed methods are benchmarked on \erdosrenyi{} graphs, $C_5$-free graphs, perturbed perfect graphs, and DIMACS instances. All benchmark instances and implementations are publicly available \citep{erdem_github_modification}.


The remainder of the paper is organized as follows. Section~\ref{sec:notation} introduces the notation and formalizes the problem definitions. Section~\ref{sec:IP} presents our integer programming formulation. Section~\ref{sec:findOddHoles} details the odd hole enumeration algorithm. In Section~\ref{sec:branch-and-cut}, we discuss the branch-and-cut framework, analyze the expected number of odd structures in random graphs, and introduce the primal heuristic. Finally, Section~\ref{sec:experiments} presents our computational results, followed by concluding remarks in Section~\ref{sec:conclusion}.

\section{Notation and Problem Definitions} \label{sec:notation}


Induced cycles of size greater than or equal to 4 are called \textit{holes}, and their complements are \textit{antiholes}. Those containing an odd number of vertices are called \textit{odd holes} and \textit{odd antiholes}, respectively. An edge between two non-consecutive vertices of a cycle is called a \textit{chord}. Accordingly, holes are cycles without chords. A graph $G$ is \textit{perfect} if for each induced subgraph $H$ of $G$, the clique number of $H$ equals the chromatic number of $H$. The Strong Perfect Graph Theorem states that a graph is perfect if and only if it does not contain odd holes or odd antiholes \citep{SPGT}.

Given a graph $G$ with vertex set $V$ and edge set $E$, we address the following graph modification problems. The \textit{minimum perfect editing problem} (\textsc{MinPerfEdit}) aims to find a perfect graph $G'=(V,E')$ that minimizes the size of the symmetric difference $E \Delta E' = (E \setminus E') \cup (E' \setminus E)$. 
The \textit{minimum perfect completion problem} (\textsc{MinPerfComp}) and the \textit{minimum perfect deletion problem} (\textsc{MinPerfDel}) restrict modifications strictly to edge additions ($E \subseteq E'$) and deletions ($E' \subseteq E$), minimizing $|E' \setminus E|$ and $|E \setminus E'|$, respectively.

The \textit{perfect sandwich problem} (\textsc{PerfSand}) is a decision problem on two input graphs $G_1 = (V, E_1)$ and $G_2 = (V, E_2)$ satisfying $E_1 \subseteq E_2$. It asks whether a perfect graph $G=(V,E)$ exists such that $E_1 \subseteq E \subseteq E_2$. Edges in $E_1$ are called \textit{mandatory edges}, while those in $E_2 \setminus E_1$ are referred to as \textit{optional edges}.

Because the empty and complete graphs are perfect, it is always possible to obtain a perfect graph via edge modifications. Thus, \textsc{MinPerfEdit}, \textsc{MinPerfComp} and \textsc{MinPerfDel} are always feasible. Conversely, \textsc{PerfSand} is a decision problem with a possibly ``no" answer. Furthermore, if $E_2$ contains all possible edges, \textsc{PerfSand} reduces to the perfect completion problem, whereas if $E_1=E_2$, it reduces to the perfect graph recognition.

\section{Integer Programming Formulation} \label{sec:IP}

To solve the aforementioned perfect modification problems, we propose an integer program that utilizes the Strong Perfect Graph Theorem. It takes any given graph as input and returns a perfect graph that is obtained with the minimum number of edge modifications. Throughout this study, we refer to the given graph as the \textit{input graph}, and the graph corresponding to the optimal (or suboptimal) solution of the integer program as the \textit{output graph}. The graph order is fixed and is denoted by $n$. The decision variables $x_{ij}$ are defined for all $(i,j)$ vertex pairs and represent the adjacency matrix entries of the output graph. That is, $x_{ij} = 1$ if $\{i,j\}$ is an edge in the output graph, and 0 otherwise. The constraints guarantee that no odd holes or odd antiholes exist in the output graph. Note that a labeled set of vertices admits several configurations, all of which induce a graph isomorphic to the ``same'' odd hole. Since the integer program formulation uses labeled vertices, the constraints need to take into account all odd hole and odd antihole configurations. Let $\mathcal{OH}$ and $\overline{\mathcal{OH}}$, respectively, denote the sets of all odd hole and odd antihole configurations on the vertex set $V$ of the input graph $G=(V, E)$. We formulate the integer program Model \texttt{IP\_Perfect} for the minimum perfect editing problem as follows.
\begin{subequations}\label{general_model}
\renewcommand{\theequation}{\theparentequation.\arabic{equation}}
\begin{alignat}{3}
	& \text{min} &\quad& \sum_{\{i,j\} \in E(G)} (1 - x_{ij}) +\sum_{\{i,j\} \in \overline{E}(G)} x_{ij} \label{general_model:obj} \\
	& \text{s.t.} && \sum_{\{i,j\} \in E(S)} (1 - x_{ij}) +  \sum_{\{i,j\} \in \overline{E}(S)} x_{ij} \geq 1 &\qquad& \forall S \in \mathcal{OH} \label{general_model:eq1} \\ 
	&&& \sum_{\{i,j\} \in E(\overline{S})} (1 - x_{ij}) +  \sum_{\{i,j\} \in \overline{E}(\overline{S})} x_{ij} \geq 1 &\qquad& \forall \overline{S} \in \overline{\mathcal{OH}} \label{general_model:eq2} \\ 
	&&& x_{ij} \in \{0,1\} && \forall \{i,j\} \in (V(G) \times V(G)), i<j 
\end{alignat}
\end{subequations}

Constraints \eqref{general_model:eq1} and \eqref{general_model:eq2} eliminate odd holes and odd antiholes, respectively. If a forbidden configuration $S$ were present, the left-hand side of its corresponding inequality would evaluate to 0, violating the constraint. Finally, the objective function \eqref{general_model:obj} minimizes the total modifications with respect to the input graph, where the first and second terms count edge removals and additions, respectively.

The sizes of the constraint sets \eqref{general_model:eq1} and \eqref{general_model:eq2} grow exponentially with respect to $n$ (see Equations~\eqref{eq:OH}~and~\eqref{eq:OHbar} in Section~\ref{sec:odd_hole_analysis}). This indicates that as the graph order increases, Model \texttt{IP\_Perfect} cannot be solved in a reasonable time. In our implementation of Model \texttt{IP\_Perfect}, we were able to solve instances up to order 10 without exceeding the computer memory. As a remedy, in Section~\ref{sec:branch-and-cut}, noting that many constraints will never be violated as the integer program is solved, we will explore the idea of dynamically generating the constraints as needed. In Sections~\ref{sec:completion}~and~\ref{sec:sandwich}, we explain how do we modify Model \texttt{IP\_Perfect} to solve \textsc{MinPerfComp} and \textsc{PerfSand}, respectively.

\section{An Algorithm for Finding Odd Holes}\label{sec:findOddHoles}

The recognition of perfect graphs can be performed in polynomial time \citep{recognizingBergeGraphs} and \citep{aPolyTimeAlgoForPerfectGraphs}. However, to implement Model \texttt{IP\_Perfect} in a branch-and-cut setting, we need to check whether the output graph is perfect and identify odd holes and odd antiholes if it is not. To this end, a recognition algorithm that merely determines whether the graph is perfect or not is insufficient. We note that polynomial-time algorithms exist for detecting the existence of an odd hole or odd antihole (in $O(n^9)$ \citep{detecting_an_odd_hole}, and $O(n^7)$ \citep{bestoddhole}). These results imply that our separation problem at integer solutions is polynomial time solvable. However, such algorithms typically return a boolean result or a single certificate, whereas in our branch-and-cut framework, each odd hole yields a distinct inequality. Preferably, we need an algorithm that identifies a batch of odd holes and odd antiholes while being time efficient in practice. As shown by our experiments in~Section~\ref{sec:edit} where we empirically evaluate the practical impact of identifying differing quantities of these structures, enumerating multiple odd holes and odd antiholes provides stronger results. Therefore, we propose a recursive backtracking algorithm, \texttt{FindOddHoles}, to enumerate all odd holes and odd antiholes; its exponential worst-case time complexity is unavoidable, as a graph can contain exponentially many such structures (see Section~\ref{sec:odd_hole_analysis}).

\vspace{0.5\baselineskip}

\begin{algorithm}[H]
	\singlespacing
	\caption{\texttt{FindOddHoles}}
	\label{algo:findOddHoles}
	
	\KwInput{A graph $G = (V(G), E(G))$}
	\KwOutput{Odd holes in the graph $G$}
		
	Let $oddHoles=\varnothing$. \\
	\For{$v \in V(G)$}{
	Let $path = \{v\}$ \\
	\texttt{FindOddHolesRecursive}($G$, $path$, $oddHoles$) \\
	}
	Return $oddHoles$
\end{algorithm}

\begin{algorithm}[H]
	\singlespacing
	\caption{\texttt{FindOddHolesRecursive}}
	\label{algo:findOddHolesRecursive}
	
	\KwInput{A graph $G$, $path$, $oddHoles$}
	
	$u\leftarrow $ The first vertex of $path$ \\
	$v\leftarrow $ The last vertex of $path$ \\
	\For{$x \in N(v), x \notin V(path)$}{\label{line:repeat_for_neighbors_of_v}
	
		\lIf(\tcp*[f]{avoid symmetrical solutions}){$x < u$}{\KwContinue}\label{line:index_comparison}
		$noChord\leftarrow \KwTrue$ \\
		\For(\tcp*[f]{chord check}){$y \in V(path)\setminus\{u,v\} $}{ \label{line:chord_check}
			\If{$(x,y) \in E(G)$}{
				$noChord \leftarrow \KwFalse$ \\
				\KwBreak \label{line:chord_check_end}
			}
		}

		\uIf(\tcp*[f]{cycle check}) {$noChord$ \KwAnd $|V(path)| \geq 2$ \KwAnd $(u,x) \in E(G)$}{ \label{line:cycle_check}
			\If (\tcp*[f]{odd hole check}){$|V(path)| \geq 4$ \KwAnd $|V(path)|$ even}{\label{line:odd_hole_check}
				Add $path \cup \{x\}$ to $oddHoles$\label{line:add_hole_to_odd_holes}
			}
		}
		\ElseIf{$noChord$}{
			Append $x$ to $path$.\label{line:append_x}\\
			\texttt{FindOddHolesRecursive}($G$, $path$, $oddHoles$)\label{line:continue_recursion}
		}
	}
	Remove $v$ from $path$ \label{line:remove_v}\\
	Return \label{line:return_recursion}
	
\end{algorithm}

\vspace{0.5\baselineskip}

Algorithm \texttt{FindOddHoles} identifies and outputs either all or, if desired, a subset of the odd holes present in a graph. To find the odd antiholes in a graph, the algorithm is applied to the complement of the graph. The outputs, which are the odd holes in the complement graph, correspond to odd antiholes in the original graph. The algorithm works by calling the recursive function with a path consisting of a single vertex, and repeats this process for every vertex of the graph. This path will be completed to the odd holes that include the first vertex ($u$ in Figure~\ref{fig:foh}) of the path, if there are any. 

\begin{figure}[H]
\centering
\resizebox{!}{5cm}{
    \begin{tikzpicture}
    
    \node[] at (1, 2.5) {\scriptsize First vertex in $path$};
    \node[draw, circle, inner sep=2pt] (u) at (2,2) {$u$};
    \node[] at (3.5, 1.4) {\scriptsize $path$};
    \node[draw, circle, inner sep=2pt] (v) at (4,0) {$v$};
    \node[] at (5.65, 0) {\scriptsize Last vertex in $path$};
    \draw[decorate, decoration={snake, amplitude=1mm, segment length=5mm}] (u) -- (v);
    \node[draw, circle, inner sep=2pt] (x) at (2.75,-0.5) {$x$};
    \draw (x) -- (v);
    \draw (x) to[out=135, in=270] (u);
    \node[] at (1.8,0.5) {\scriptsize ?};
    \draw[loosely dashed] (x) to[out=105, in=255] (2.4, 1.2);
    \draw[loosely dashed] (x) to[out=75, in=255] (3.2, 0.3);
    \node[] at (2.8,0.5) {\scriptsize ?};
    
    \end{tikzpicture}
}
\caption{An illustration of Algorithm \texttt{FindOddHoles} conducting an odd hole check.}
\label{fig:foh}
\end{figure}

Inside the recursive function, paths emerging from the first vertex are created recursively. The crucial part of the algorithm is that before a neighbor ($x$) of the last vertex ($v$) of the path is appended to the path, a check is carried out for existence of a hole. First, if the candidate vertex has any neighbors within the internal vertices (vertices not including the first and the last) of the path, then this candidate vertex is not considered further since such an edge would correspond to a chord in the potential cycle. If there is no chord, then the adjacency of the candidate vertex to the first vertex in the path is checked. If they are not adjacent, then there is no cycle in the subgraph induced by the path and the candidate vertex. Then the candidate vertex is added to the end of the path and the recursive function is called with this new path. If the candidate vertex is adjacent to the first vertex of the path, then the path together with the candidate vertex is a chordless cycle. If this chordless cycle is of odd length greater than or equal to 5, it is an odd hole. This odd hole is added to a data structure to keep a record. This process is repeated with all the neighbors of the last vertex of the path. When the neighbors are depleted, the last vertex is removed from the path and the recursive function returns.


\section{Branch-and-Cut Algorithm} \label{sec:branch-and-cut}

In this section, we present a branch-and-cut algorithm that dynamically generates constraints to obtain a perfect graph with the minimum number of modifications to the input graph. Based on Model \texttt{IP\_Perfect} presented in Section~\ref{sec:IP}, our approach initializes the model with only the odd holes and odd antiholes present in the input graph, rather than all possible odd hole and odd antihole configurations. During the exploration of the branch-and-cut tree, we encounter integer solutions, which we call \textit{candidate graphs}. Whenever a candidate graph is found, Algorithm \texttt{FindOddHoles} is executed on it and its complement. If no odd holes or odd antiholes are found, the graph is perfect; thus, it is a feasible solution to the problem, providing an upper bound. Otherwise, the graph is not perfect, and we add cuts corresponding to the detected structures as lazy constraints. These cuts eliminate the infeasible integer point, allowing the solver to resume its search.




The polyhedral characterization of perfect graphs given by \citet{chvatal75} implies that a graph $G$ is perfect if and only if its stable-set polytope STAB($G$) coincides with its 
fractional relaxation defined by clique inequalities QSTAB($G$). The Strong Perfect Graph Theorem characterization of perfect graphs implies that odd holes and odd antiholes give rise to fractional extreme points of QSTAB($G$). Although our formulation does not explicitly introduce stable-set or clique variables, the inequalities generated by our branch-and-cut algorithm can be interpreted as eliminating solutions that lead to fractional extreme points of QSTAB($G$). From this perspective, our branch-and-cut algorithm progressively restricts the space of admissible graphs until the fractional relaxation of the stable-set polytope QSTAB($G$) associated with the resulting graph is integral, at which point the graph is perfect by the Strong Perfect Graph Theorem.

\begin{algorithm}[!t]
    \vspace{-0.5\baselineskip}
	\singlespacing
	\caption{\texttt{Branch-and-Cut}} 
	\label{algo:BaC}
	\KwInput{A graph $G = (V(G), E(G))$}
	\KwOutput{A perfect graph obtained by minimum edge modifications on $G$}
	
	Let $\widehat{OH} \leftarrow \texttt{FindOddHoles}(G)$. \label{line:init_oh} \\
	Let $\widehat{\overline{OH}} \leftarrow \texttt{FindOddHoles}(\overline{G})$.  \label{line:init_oah} \\
	Initialize Model \texttt{IP\_Perfect} using $G$, $\widehat{OH}$ and $\widehat{\overline{OH}}$. \\
	
	\BlankLine
	\textbf{Define} Lazy Constraint Callback for intermediate integer solution $G_x$: \\
	\Indp
		$newOH \leftarrow \texttt{FindOddHoles}(G_x)$. \label{line:foh1} \\
		$newOAH \leftarrow \texttt{FindOddHoles}(\overline{G_x})$. \label{line:foh2} \\
		\If{$newOH \neq \varnothing$ \KwOr $newOAH \neq \varnothing$}{
			Add $newOH$ and $newOAH$ to \texttt{IP\_Perfect} as lazy cuts .\label{line:add_lazy_cuts}
		}
	\Indm
	\BlankLine
	
	Solve Model \texttt{IP\_Perfect} using the Lazy Constraint Callback. \label{line:solve_ip_in_bc} \\
	Let $G^x$ be the solution returned by the solver. \\
	Return $G^*$.
\end{algorithm}


To accelerate this procedure in practice, we do not need to identify every odd hole and odd antihole to invalidate an imperfect candidate graph; generating a subset of violated cuts is sufficient to eliminate the current integer solution. To exploit this, we introduce a new parameter, \texttt{ODD\_HOLE\_TERMINATION\_PERCENTAGE} (\texttt{OHTP}). For \erdosrenyi~graphs, \texttt{OHTP} is multiplied by the expected number of odd holes to define a dynamic termination threshold for Algorithm \texttt{FindOddHoles}. Unlike a fixed numerical limit, this adaptive threshold automatically scales with varying graph orders and densities. We apply the exact same thresholding approach to odd antiholes, utilizing their respective expected numbers.

\subsection{The Number of Odd Holes and Odd Antiholes} \label{sec:odd_hole_analysis}

In our experiments, an important set of input graphs is random \erdosrenyi{} graphs $G_{ER(n,p)}$, where $n$ is the number of vertices and $p$ is the edge probability. Accordingly, we investigate the expected number of odd holes and odd antiholes in Erd\H{o}s--R\'enyi graphs. Although some basic counting and probability knowledge suffices to derive the expected number of odd holes and odd antiholes in Erd\H{o}s--R\'enyi graphs, to the best of our knowledge, this number has not been investigated in the literature. Therefore, we provide full analysis for the sake of completeness.


\begin{thm}\label{thm:expectednumbers}
    The expected number of odd holes and odd antiholes in \erdosrenyi{} graphs are given in Equations~\eqref{eq:E_X}~and~\eqref{eq:E_Xbar}, respectively.
\end{thm}

\begin{equation}
    \mathbb{E}[X] = \frac{n!}{2} \sum_{k = 2}^{\left\lfloor \frac{n-1}{2} \right\rfloor} \frac{p^{2k+1} (1-p)^{(k-1)(2k+1)}}{(2k+1)(n-2k-1)!} .
    \label{eq:E_X}
\end{equation}
\begin{equation}
\mathbb{E}[\overline{X}] = \frac{n!}{2} \sum_{k = 3}^{\left\lfloor \frac{n-1}{2} \right\rfloor} \frac{p^{(2k+1)(k-1)}(1-p)^{(2k+1)}}{(2k+1)(n-2k-1)!}.
\label{eq:E_Xbar}
\end{equation}

\begin{proof}
First, let us count the number of odd hole and odd antihole configurations on $n$ vertices. There are $\frac{(i-1)!}{2}$ odd hole configurations on $i$ vertices. Let us define the set of all odd hole configurations as $\mathcal{OH}$ and the set of odd hole configurations of length $i$ as $\mathcal{OH}_{i}$. In Equations \eqref{eq:OH_i} and \eqref{eq:OH}, we derive the numbers of odd hole configurations of length $i$ and all odd hole configurations on $n$ vertices, respectively.

\begin{equation}
\label{eq:OH_i}
| \mathcal{OH}_i | = {n \choose i} \frac{(i-1)!}{2} = \frac{n!}{2i(n-i)!}
\end{equation}

\begin{equation}
\label{eq:OH}
\mathcal{|OH|} =  \sum_{k = 2}^{\left\lfloor \frac{n-1}{2} \right\rfloor} {n \choose 2k+1} \frac{(2k)!}{2} = \frac{n!}{2} \sum_{k = 2}^{\left\lfloor \frac{n-1}{2} \right\rfloor} \frac{1}{(2k+1)(n-2k-1)!}
\end{equation}

The number of odd antihole configurations is calculated in a similar way. An odd hole's complement is an odd antihole. 
Since a $C_5$ is self complementary, meaning that its complement is also a $C_5$, it is reasonable to skip $C_5$ cycles when counting the number of odd antiholes. The set of odd antihole configurations is denoted by $\overline{\mathcal{OH}}$ and its size in a graph of order $n$ is given in \eqref{eq:OHbar}.

\begin{equation}
\label{eq:OHbar}
|\overline{\mathcal{OH}}| = \sum_{k = 3}^{\left\lfloor \frac{n-1}{2} \right\rfloor} {n \choose 2k+1} \frac{(2k)!}{2} = \frac{n!}{2} \sum_{k = 3}^{\left\lfloor \frac{n-1}{2} \right\rfloor} \frac{1}{(2k+1)(n-2k-1)!}
\end{equation}

Equations \eqref{eq:OH} and \eqref{eq:OHbar} give the cardinalities of the constraint sets \eqref{general_model:eq1} and \eqref{general_model:eq2} of Model \texttt{IP\_Perfect}, respectively.

We now count the number of odd holes of a given graph. Given a graph $G$, the indicator function $\mathbb{I}_{G}(j): \mathcal{OH} \to \{0,1\}$ is defined as
\begin{equation}
\mathbb{I}_{G}(j) = 
\begin{cases} 
1 & \text{if odd hole configuration } j \text{ is realized in } G,\\
0 & \text{otherwise}.
\end{cases}
\end{equation}

The number of odd holes of size $2k+1$ in a given graph $G$ is denoted by $X_{2k+1}$ and the total number of odd holes in $G$ is denoted by $X$. Clearly,
\begin{equation}
\label{eq:n_of_odd_holes}
X = \sum_{k = 2}^{\left\lfloor \frac{n-1}{2} \right\rfloor} X_{2k+1} =  \sum_{j \in \mathcal{OH}} \mathbb{I}_{G}(j) = \sum_{k = 2}^{\left\lfloor \frac{n-1}{2} \right\rfloor} \sum_{j \in \mathcal{OH}_{2k+1}} \mathbb{I}_{G}(j)
\end{equation}

In a given graph, $X_{2k+1}$ and $X$ can be calculated by counting. However, for a random Erd\H{o}s--R\'enyi graph, $G_{ER(n,p)}$, $X_{2k+1}$ and $X$ are random variables. Moreover, the occurrence of an odd hole with configuration $j$ is a Bernoulli trial with probability mass function
\begin{equation}
\label{eq:bernoulli}
P(\mathbb{I}_{G_{ER(n,p)}}(j) = x) = 
\begin{cases} 
p^{|j|} (1-p)^{{|j| \choose 2} - |j|} & \text{if } x = 1 \\
1 - (p^{|j|} (1-p)^{{|j| \choose 2} - |j|}) & \text{if } x = 0,
\end{cases} 
\end{equation}
where $|j|$ is the length of the odd hole configuration.

The number of odd holes in a random Erd\H{o}s--R\'enyi graph is a random variable. We are also interested in its sample space (the set of values it can take). This random variable takes non-negative integer values and the lowest value it can take is 0. The highest value it can take is presented in \citep{max_induced_cycles}. In \citep{lidicky_max_c5}, extremal graphs of given order which have the maximum number of induced $C_5$ cycles are constructed.

Additionally, to calculate the expected number of odd holes in a random \erdosrenyi{} graph, we build on Equation \eqref{eq:n_of_odd_holes} using the linearity of the expectation operator. 
\begin{equation}\label{eq:E_X_1}
\mathbb{E}[X] = \sum_{k = 2}^{\left\lfloor \frac{n-1}{2} \right\rfloor} \sum_{j \in \mathcal{OH}_{2k+1}} \mathbb{E}[ \mathbb{I}_{G}(j) ] = \sum_{k = 2}^{\left\lfloor \frac{n-1}{2} \right\rfloor} \sum_{j \in \mathcal{OH}_{2k+1}} p^{|j|} (1-p)^{{|j| \choose 2} - |j|}
\end{equation}
The expectation of a Bernoulli trial is simply the probability (given in Equation \eqref{eq:bernoulli}) that the trial takes the value 1. In one term of the outer sum where $k$ is constant, the length of the odd hole configuration $j$ can be expressed in terms of $k$ and is equal to $2k+1$. With the same reasoning, the inner sum can be represented as summing over the size of the set $\mathcal{OH}_{2k+1}$.
Clearly, the inner sum is the summation of the same terms for $| \mathcal{OH}_{2k+1} |$ many times. Moreover, the size of this set can be expressed with $n$ and $k$ as noted in Equation \eqref{eq:OH_i}. 
\begin{equation}
\mathbb{E}[X] = \sum_{k = 2}^{\left\lfloor \frac{n-1}{2} \right\rfloor} {n \choose {2k+1}} \frac{(2k+1-1)!}{2} p^{2k+1} (1-p)^{{2k+1 \choose 2} - (2k+1)}
\end{equation}
By expanding the combination and simplifying the terms, we obtain the expected number of odd holes given in Equation~\eqref{eq:E_X}.

The total number of odd antiholes can be calculated similarly. Denote the number of odd antiholes in a random \erdosrenyi{} graph by $\overline{X}$. Then, the expectation of $\overline{X}$ is given in Equation \eqref{eq:E_X_bar_1} and has little alteration of power terms of $p$ and $1-p$ in Equation \eqref{eq:E_X_1}. As mentioned above, $C_5$ is self-complementary and they are included in the expected number of odd holes. It is reasonable to omit $C_5$ cycles in the calculation of expected value of odd antiholes to avoid double counting. Thus, we consider the antiholes of size 7 and higher.
\begin{equation}
\label{eq:E_X_bar_1}
\mathbb{E}[\overline{X}] = \sum_{k = 3}^{\left\lfloor \frac{n-1}{2} \right\rfloor} \sum_{j \in \overline{\mathcal{OH}}_{2k+1}} p^{{|j| \choose 2} - |j|} (1-p)^{|j|}
\end{equation}
We simplify Equation \eqref{eq:E_X_bar_1} using Equation \eqref{eq:OHbar} and obtain the expectation of odd antiholes as Equation~\eqref{eq:E_Xbar}.
\end{proof}

\begin{figure}[H]
	\centering
    
    \begin{subfigure}{0.47\textwidth}
    	\centering
    	\includegraphics[width=\textwidth]{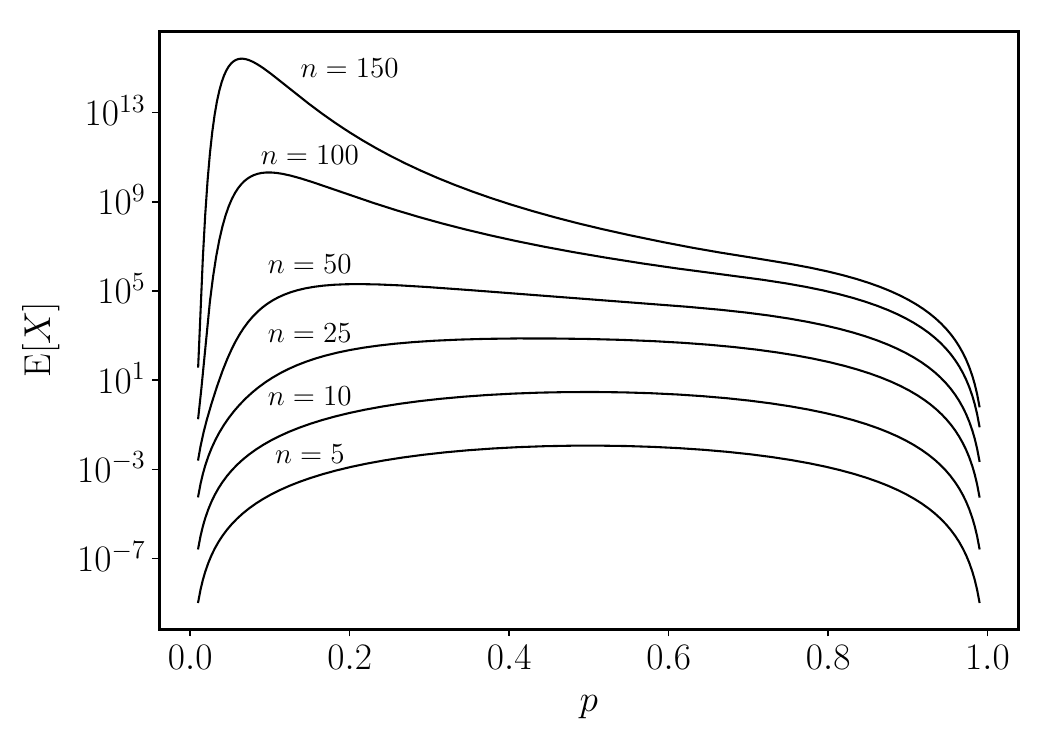}
    	\caption{$\mathbb{E}[X]$}
    	\label{fig:E_X_on_p_line_for_n}
    \end{subfigure}
    \hspace{0.25\baselineskip}
    \begin{subfigure}{0.47\textwidth}
    	\includegraphics[width=\textwidth]{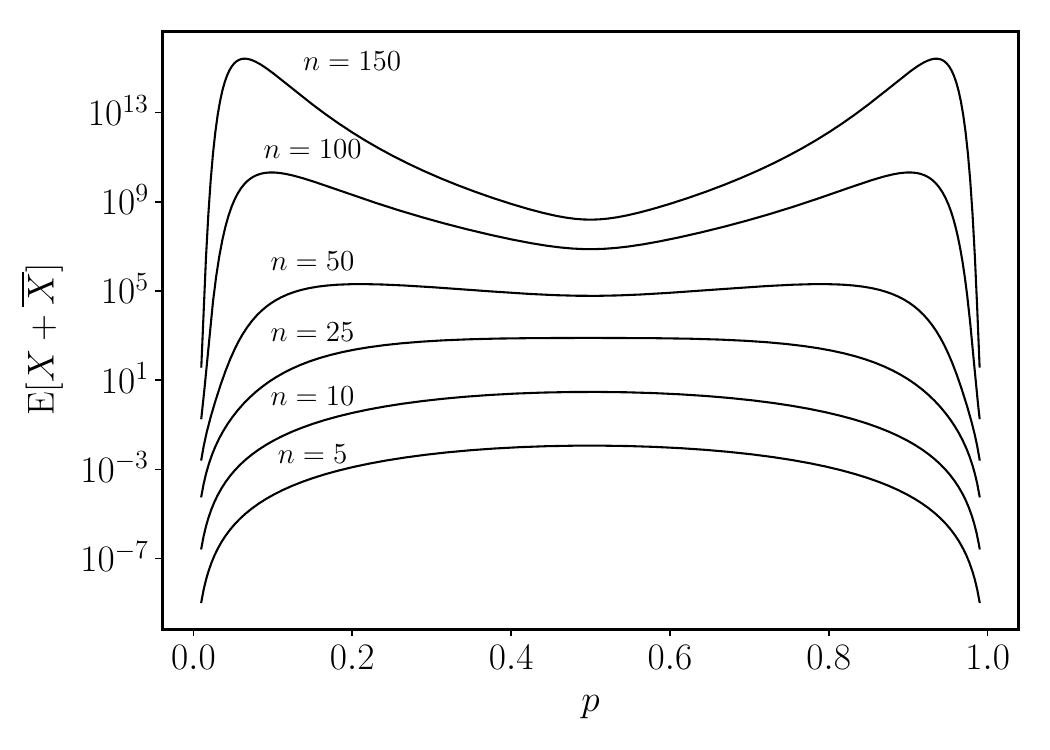}
    	\caption{$\mathbb{E}[X + \overline{X}]$}
    	\label{fig:E_X_Xbar_on_p_line_for_n}
    \end{subfigure}

    \caption{The expected number of odd holes (and odd antiholes) in random \erdosrenyi{} graphs for several graph orders (in logarithmic $y$-scale).}
    
\end{figure}

Figure \ref{fig:E_X_on_p_line_for_n} illustrates the expected number of odd holes as a function of graph order and edge probability $p$. As the graph order increases, the maximizing density shifts from $0.5$ toward $0$. This occurs because odd holes of size greater than $5$ contain more non-edges than edges. Similarly, Figure \ref{fig:E_X_Xbar_on_p_line_for_n} displays the combined expected number of odd holes and odd antiholes, which is strictly symmetric around $p = 0.5$. For graphs of low order, the expected value has a single peak at $p = 0.5$. However, as the graph order increases, two distinct peaks emerge and migrate outward to densities $0$ and $1$.

In our experiments (presented in Section \ref{sec:experiments}), we use the expected numbers of odd holes and odd antiholes of the \erdosrenyi{} graphs to decide which percentage of these numbers should be added as constraints at every iteration of the branch-and-cut algorithm.

\subsection{A Primal Heuristic for Perfect Graph Modification} \label{sec:imh}

Within our branch-and-cut framework, the upper bound is updated whenever a feasible solution---which, in our context, is a perfect graph---is discovered. Identifying high-quality feasible solutions tightens the upper bound and accelerates the pruning of the search tree, thereby improving overall algorithmic performance. For this purpose, we propose a primal heuristic called \texttt{IterativeModificationHeuristic} (\texttt{IMH}) for \textsc{MinPerfEdit} (its adaptation for \textsc{MinPerfComp} is detailed in Section~\ref{sec:completion}).

The \texttt{IMH} is a greedy local search algorithm that iteratively reduces the number of odd holes and odd antiholes by modifying edges until none remain, thereby yielding a perfect graph. Initially, the algorithm enumerates all such structures present in the input graph. At each iteration, it identifies the vertex pair included in the maximum number of odd holes and odd antiholes and toggles its adjacency state (i.e., an edge is removed or a non-edge is added). While toggling this state destroys all existing violations that include the vertex pair, it may also introduce new ones. However, because any newly created structures must inherently contain the modified pair, we avoid re-evaluating the entire graph from scratch. Instead, a localized variant of \texttt{FindOddHoles} efficiently searches only for new structures containing this specific pair. If the number of newly created structures meets or exceeds the number of eliminated ones, the toggle is reverted, and the vertex pair is flagged as failed for the current iteration. Otherwise, the algorithm accepts the modification and proceeds to the next iteration.

By repeating this process, the total number of odd holes and odd antiholes decreases each iteration, eventually yielding a perfect graph—unless the algorithm reaches a state where no vertex-pair modification reduces this total. This rarely arose in our experiments but its occurrence is not detrimental, since the heuristic's primary goal is merely to find feasible solutions giving better upper bounds. However, such cases demonstrate that some graphs admit no edge modification that strictly decreases the total number of odd holes and odd antiholes. For these graphs, the \texttt{IMH} cannot find a perfect graph, since it does not allow temporary increases in this total.

\begin{algorithm}[!t]
	\singlespacing
	\caption{\texttt{IterativeModificationHeuristic (IMH)}} 
	\label{algo:heuristic}
	
	\KwInput{A candidate graph $G_c$}
	\KwOutput{A perfect graph if successful}
	
	$OH \leftarrow \texttt{FindOddHoles}(G_c)$. \\
	$OAH \leftarrow \texttt{FindOddHoles}(\overline{G_c})$. \\
	$vpCounts[vp] \leftarrow |\{H \in OH \cup OAH | vp \in H\}|$, \textbf{foreach} $vp$.\tcp*[f]{$vp$: vertex pair} \\
	\While{$|OH| + |OAH| > 0$}{
		Let $failedVertexPairs \leftarrow \varnothing$. \\
		\While{\KwTrue}{
			\If(\tcp*[f]{Heuristic failure check.}){$\forall vp \in failedVertexPairs$}{
				Return as failed.
			}
			Let $vp^* \leftarrow \displaystyle \arg\max_{\substack{vp \notin \textit{failedVertexPairs}}} \textit{vpCounts}[vp]$.\label{line:heur_identify_vp} \\
			Change the state of $G_c[vp^*]$. \label{line:heur_change_vp} \\
			$newOH \leftarrow \texttt{FindOddHoles\_v2}(G_c, vp^*)$. \label{findOddHolesMod1} \\
			$newOAH \leftarrow \texttt{FindOddHoles\_v2}(\overline{G_c}, vp^*)$. \label{findOddHolesMod2} \\
			\If(\tcp*[f]{Revert the change.}){$|newOH| + |newOAH| \geq vpCounts[vp^*]$}{ \label{line:heur_revert_start}
				Change the state of $G_c[vp^*]$. \\
				Add $vp^*$ to $failedVertexPairs$. \label{line:heur_revert_end} \\
			}
			\Else(\tcp*[f]{Continue with the change.}){ \label{line:heur_continue_start} 
				Remove holes containing $vp^*$ from $OH$ and $OAH$. \\
				Add $newOH$ into $OH$, add $newOAH$ into $OAH$. \\
                $failedVertexPairs \leftarrow \varnothing$. \\
				Recalculate $vpCounts$. \\
				\KwBreak \label{line:heur_continue_end}
			}
		}
	}
	Return $G_c$
\end{algorithm}

Algorithm \texttt{IMH} operates deterministically, producing consistent output for the same input. This is not desirable and some variation is needed since the goal is to improve the upper bounds. Two options to overcome this situation are to introduce a stochastic selection criterion within the heuristic or to initiate the heuristic with different inputs. The latter option fits perfectly into the solution scheme of the branch-and-cut algorithm. We choose to pass each candidate graph generated in the branch-and-cut algorithm as an input to the heuristic. Furthermore, we obtain graphs from the fractional solutions of the relaxation nodes of the branch-and-cut tree (see Section \ref{sec:experiments} for details).

\section{Experimental Results} \label{sec:experiments}

In this section, we evaluate the performance of our branch-and-cut algorithm on \textsc{MinPerfEdit} (Section~\ref{sec:edit}), \textsc{MinPerfComp} (Section~\ref{sec:completion}) and \textsc{PerfSand} (Section~\ref{sec:sandwich}). Moreover, we assess the performance of the \texttt{IMH} as a standalone algorithm in Section~\ref{sec:exp_imh}.

All algorithms were implemented in C++20, and the integer programming framework was solved using IBM ILOG CPLEX version 22.1.1.0. We conducted the experiments on a machine equipped with 64 GB of RAM and a 12th Gen Intel Core i9-12900 CPU (utilizing 24 threads). We impose a strict runtime limit of 15 minutes per instance; accordingly, the algorithm terminates either upon finding an optimal solution or reaching this time limit.

To analyze algorithmic performance across diverse structural properties, we conduct experiments on a wide range of instance types, including random \erdosrenyi{}, $C_5$-free, and perturbed perfect graphs, as well as several DIMACS benchmark instances. With the exception of the DIMACS instances, we generated five distinct graphs for each graph type, order $n$, and target density $d \in \{0.25, 0.50, 0.75\}$.

Motivated by the finding that almost all $C_5$-free graphs are perfect \citep{c5free}, we investigate whether perfect graph modification problems can be solved more efficiently on these graphs in practice. To generate a random $C_5$-free graph, we initialize an empty graph on $n$ vertices and iteratively add random edges until the target density is reached, ensuring no added edge creates an induced $C_5$. Since this scheme occasionally stalls, where any edge addition would induce a $C_5$, such graphs are discarded and generation restarts. As target density increases, stalling grows more frequent. Thus, to generate instances with density 0.75, we generated $C_5$-free graphs with density 0.25 and took their complements. Since the complement of a $C_5$-free graph is also $C_5$-free, this approach is mathematically sound. Finally, we verify perfectness of each graph using \texttt{is\_perfect} \citep{erdem_is_perfect}; if already perfect, it is discarded and a new one generated.

We generated another set of instances that are, in some sense, measurably closer to perfect graphs. We achieve this by first generating a perfect graph using the algorithm developed by \citet{SEKER202167}, which combine small perfect graphs to construct larger ones. For the initial pool of small perfect graphs, we utilized the perfect graph database compiled by \citet{mckay_perfect_graphs}. Once a large perfect graph is constructed, we perturb it to destroy its perfectness by toggling the adjacency states of a subset of vertex pairs. The number of perturbations is calculated as a ratio of the total number of vertex pairs, and the pairs to be perturbed are selected uniformly at random. We utilized perturbation ratios of 0.1 and 0.3 in our experiments. Note that for these instances, the initial unperturbed perfect graph provides a valid upper bound for the perfect editing problem.

We use the same graphs as input for both \textsc{MinPerfEdit} and \textsc{MinPerfComp}. This allows for a direct comparison of the practical difficulty between the two problems. For \textsc{PerfSand}, we present results only on \erdosrenyi{} graphs, since the instances require an additional step of generating optional edges, making the initial graph type significantly less critical than in the editing and completion problems. The generation of the optional edges for \textsc{PerfSand} are explained in Section \ref{sec:sandwich}.

The implementations of our algorithms are available online along with the graphs used~\citep{erdem_github_modification}. The graph generation algorithms are also available for use if one wishes to generate similar types of graphs for their research~\citep{erdem_github_generation}. Moreover, we implemented Algorithm \texttt{FindOddHoles} as a standalone program named \texttt{is\_perfect} for perfect graph recognition. It is available online, accompanied with our report on its average runtime across varying graph types, orders, and densities~\citep{erdem_is_perfect}.

\subsection{The Minimum Perfect Editing Problem} \label{sec:edit}

In this section, we evaluate the empirical performance of our exact methods on \textsc{MinPerfEdit}. We begin by analyzing the impact of the \texttt{OHTP} parameter on \erdosrenyi{} graphs. At the most restrictive extreme, the separation routine terminates immediately after identifying a single violated odd hole or odd antihole. At the opposite extreme, the algorithm exhaustively enumerates all existing violations in the candidate graph. Between these boundaries, we test \texttt{OHTP} settings of 5\%, 25\%, and 50\%. For a given graph, this percentage is multiplied by the expected number of odd holes and odd antiholes to dynamically calculate a termination threshold for Algorithm \texttt{FindOddHoles}.

Table~\ref{experiment_edit_ohtp_er} reports the results of this experiment, where each row aggregates the outcomes of five distinct instances of identical order and density. The `Solved' column indicates the number of instances (out of five) solved to proven optimality, while the `Time(s)' column denotes the average runtime in seconds. The `Gap(\%)' column reports the average final optimality gap, calculated as $\frac{\mathrm{UB} - \mathrm{LB}}{\mathrm{UB}} \times 100$ at the termination of the branch-and-cut algorithm. The final row displays the overall mean values across all instances. Based on these averages, an \texttt{OHTP} of 5\% appears to be marginally the most effective; however, the performance differences across the tested thresholds are practically negligible. 

\begin{table}[!htbp]
\centering
\setlength{\tabcolsep}{4pt}
\caption{Experimental results for the \texttt{ODD\_HOLE\_TERMINATION\_PERCENTAGE} parameter in the minimum perfect editing problem on \erdosrenyi{} graphs.}
\label{experiment_edit_ohtp_er}
\resizebox{0.9\textwidth}{!}{
\begin{tabular}{cc|rrr|rrr|rrr|rrr|rrr}
\toprule
  &   & \multicolumn{15}{c}{ODD HOLE TERMINATION PERCENTAGE} \\
  &   & \multicolumn{3}{c}{One} & \multicolumn{3}{c}{5\%} & \multicolumn{3}{c}{25\%} & \multicolumn{3}{c}{50\%} & \multicolumn{3}{c}{All} \\
 $n$ & $d$ & Solved & Gap\% & Time(s) & Solved & Gap\% & Time(s) & Solved & Gap\% & Time(s) & Solved & Gap\% & Time(s) & Solved & Gap\% & Time(s) \\
\midrule
\multirow[t]{3}{*}{20} & 0.25 & 5 & 0.0 & 0 & 5 & 0.0 & 0 & 5 & 0.0 & 0 & 5 & 0.0 & 0 & 5 & 0.0 & 0 \\
  & 0.50 & 5 & 0.0 & 0 & 5 & 0.0 & 0 & 5 & 0.0 & 0 & 5 & 0.0 & 0 & 5 & 0.0 & 1 \\
  & 0.75 & 5 & 0.0 & 0 & 5 & 0.0 & 0 & 5 & 0.0 & 0 & 5 & 0.0 & 0 & 5 & 0.0 & 0 \\
\cline{1-17}
\multirow[t]{3}{*}{25} & 0.25 & 5 & 0.0 & 1 & 5 & 0.0 & 1 & 5 & 0.0 & 1 & 5 & 0.0 & 1 & 5 & 0.0 & 1 \\
  & 0.50 & 5 & 0.0 & 28 & 5 & 0.0 & 34 & 5 & 0.0 & 30 & 5 & 0.0 & 30 & 5 & 0.0 & 24 \\
  & 0.75 & 5 & 0.0 & 1 & 5 & 0.0 & 1 & 5 & 0.0 & 1 & 5 & 0.0 & 1 & 5 & 0.0 & 1 \\
\cline{1-17}
\multirow[t]{3}{*}{30} & 0.25 & 4 & 1.4 & 185 & 5 & 0.0 & 56 & 5 & 0.0 & 56 & 5 & 0.0 & 56 & 5 & 0.0 & 56 \\
  & 0.50 & 0 & 18.4 & 901 & 0 & 19.4 & 901 & 0 & 16.6 & 901 & 0 & 17.7 & 901 & 0 & 15.4 & 901 \\
  & 0.75 & 5 & 0.0 & 308 & 5 & 0.0 & 315 & 4 & 1.0 & 383 & 3 & 3.0 & 514 & 4 & 1.0 & 555 \\
\cline{1-17}
\multirow[t]{3}{*}{35} & 0.25 & 0 & 19.6 & 901 & 0 & 17.2 & 901 & 0 & 17.5 & 901 & 0 & 17.2 & 901 & 0 & 17.2 & 901 \\
  & 0.50 & 0 & 35.3 & 901 & 0 & 32.9 & 900 & 0 & 32.6 & 901 & 0 & 32.5 & 900 & 0 & 33.8 & 901 \\
  & 0.75 & 0 & 27.7 & 901 & 0 & 13.2 & 901 & 0 & 13.2 & 901 & 0 & 14.9 & 901 & 0 & 12.6 & 901 \\
\cline{1-17}
\multirow[t]{3}{*}{40} & 0.25 & 0 & 67.1 & 901 & 0 & 58.7 & 901 & 0 & 58.7 & 901 & 0 & 77.3 & 901 & 0 & 58.7 & 901 \\
  & 0.50 & 0 & 48.7 & 900 & 0 & 44.5 & 900 & 0 & 51.3 & 900 & 0 & 58.0 & 900 & 0 & 66.3 & 900 \\
  & 0.75 & 0 & 47.2 & 901 & 0 & 33.9 & 901 & 0 & 31.7 & 901 & 0 & 38.6 & 901 & 0 & 45.3 & 901 \\
\cline{1-17}
  & Mean & 2.60 & 17.7 & 455 & 2.67 & 14.6 & 448 & 2.60 & 14.8 & 452 & 2.53 & 17.3 & 460 & 2.60 & 16.7 & 463 \\
\cline{1-17}
\bottomrule
\end{tabular}
}
\end{table}

Furthermore, as shown in Table~\ref{experiment_edit_ohtp_er_foh} the time consumed by the \texttt{FindOddHoles} function is negligible compared to the total runtime: less than 1 second compared to approximately 455 seconds, as established in Table~\ref{experiment_edit_ohtp_er}. Because the optimality gaps are comparable, and because we lack explicit formulas for the expected number of violations in instances other than \erdosrenyi{} graphs, we ultimately choose the \texttt{All} option. By configuring the \texttt{FindOddHoles} function to exhaustively search for all odd holes and odd antiholes in every candidate graph, we ensure that our methodology remains uniform and directly comparable across all graph types in subsequent experiments.


\begin{table}[!htbp]
\centering
\setlength{\tabcolsep}{4pt}
\caption{Average runtime and call-count of Algorithm \texttt{FindOddHoles} for the experiment presented in Table~\ref{experiment_edit_ohtp_er}.}
\label{experiment_edit_ohtp_er_foh}
\resizebox{0.8\textwidth}{!}{
\begin{tabular}{cc|rr|rr|rr|rr|rr}
\toprule
  &   & \multicolumn{10}{c}{ODD HOLE TERMINATION PERCENTAGE} \\
  &   & \multicolumn{2}{c}{One} & \multicolumn{2}{c}{5\%} & \multicolumn{2}{c}{25\%} & \multicolumn{2}{c}{50\%} & \multicolumn{2}{c}{All} \\
$n$ & $d$  & Time(s)& \#Calls & Time(s) & \#Calls & Time(s) & \#Calls & Time(s) & \#Calls & Time(s) & \#Calls \\
\midrule
\multirow[t]{3}{*}{20} & 0.25 & 0 & 80 & 0 & 48 & 0 & 48 & 0 & 48 & 0 & 48 \\
  & 0.50 & 0 & 126 & 0 & 94 & 0 & 100 & 0 & 81 & 0 & 88 \\
  & 0.75 & 0 & 46 & 0 & 44 & 0 & 44 & 0 & 43 & 0 & 43 \\
\cline{1-12}
\multirow[t]{3}{*}{25} & 0.25 & 0 & 276 & 0 & 99 & 0 & 99 & 0 & 99 & 0 & 99 \\
  & 0.50 & 0 & 353 & 0 & 251 & 0 & 173 & 0 & 141 & 0 & 137 \\
  & 0.75 & 0 & 226 & 0 & 133 & 0 & 112 & 0 & 127 & 0 & 104 \\
\cline{1-12}
\multirow[t]{3}{*}{30} & 0.25 & 1 & 4092 & 0 & 166 & 0 & 169 & 0 & 169 & 0 & 170 \\
  & 0.50 & 0 & 3126 & 0 & 492 & 0 & 211 & 0 & 243 & 0 & 188 \\
  & 0.75 & 0 & 848 & 0 & 216 & 0 & 168 & 0 & 231 & 0 & 260 \\
\cline{1-12}
\multirow[t]{3}{*}{35} & 0.25 & 4 & 13526 & 0 & 229 & 0 & 229 & 0 & 228 & 0 & 227 \\
  & 0.50 & 1 & 3680 & 0 & 398 & 0 & 362 & 0 & 363 & 0 & 280 \\
  & 0.75 & 2 & 10450 & 0 & 232 & 0 & 312 & 0 & 267 & 0 & 195 \\
\cline{1-12}
\multirow[t]{3}{*}{40} & 0.25 & 7 & 16025 & 1 & 256 & 1 & 255 & 8 & 16214 & 1 & 257 \\
  & 0.50 & 1 & 3708 & 0 & 260 & 0 & 163 & 0 & 120 & 0 & 112 \\
  & 0.75 & 1 & 6981 & 0 & 444 & 0 & 178 & 1 & 240 & 1 & 128 \\
\cline{1-12}
  & Mean & 1.1 & 4236.69 & 0.1 & 224.51 & 0.1 & 175.36 & 0.7 & 1241.39 & 0.2 & 156.21 \\
\cline{1-12}
\bottomrule
\end{tabular}
}
\end{table}

In the following experiments, we investigate the effects of different strategies regarding when to apply the \texttt{IMH}. Table~\ref{experiment_edit_strat_er} presents the results on \erdosrenyi{} graphs. The first set of columns under `Base' represents the baseline setting without the heuristic and corresponds to the final set of columns of Table~\ref{experiment_edit_ohtp_er}. In the second set of columns, `H on C', the \texttt{IMH} is run exclusively on integer solutions (candidate graphs). The column `H on R' indicates the application of the \texttt{IMH} at the relaxation nodes of the branch and cut tree. Because a solution at a relaxation node is fractional, we obtain a graph by rounding decision variables (vertex pairs) with values $\geq 0.5$ up to 1, and rounding the remaining variables down to 0. We observed that repeating this method at every relaxation node of the branch-and-cut tree over-allocates time to the heuristic. Thus, under this strategy, the heuristic is called only at every tenth node. Lastly, the fourth set of columns `H on C\&R'---which represents the strategy of combining the two methods---yields the best overall performance compared to the alternatives.

\begin{table}
\centering
\setlength{\tabcolsep}{4pt}
\caption{Experimental results for the minimum perfect editing problem on Erd\H{o}s--R\'enyi graphs using various strategy settings of \texttt{IterativeModificationHeuristic}.}
\label{experiment_edit_strat_er}
\resizebox{0.9\textwidth}{!}{
\begin{tabular}{cc|rrr|rrr|rrr|rrr|rrr}
\toprule
 &  & \multicolumn{12}{c}{Strategies} \\
 &  & \multicolumn{3}{c}{Base} & \multicolumn{3}{c}{H on C} & \multicolumn{3}{c}{H on R} & \multicolumn{3}{c}{H on C\&R} \\
 $n$ & $d$ & Solved & Gap\% & Time(s) & Solved & Gap\% & Time(s) & Solved & Gap\% & Time(s) & Solved & Gap\% & Time(s) \\
\midrule
\multirow[t]{3}{*}{20} & 0.25 & 5 & 0.0 & 0 & 5 & 0.0 & 0 & 5 & 0.0 & 0 & 5 & 0.0 & 0 \\
 & 0.50 & 5 & 0.0 & 1 & 5 & 0.0 & 0 & 5 & 0.0 & 0 & 5 & 0.0 & 0 \\
 & 0.75 & 5 & 0.0 & 0 & 5 & 0.0 & 0 & 5 & 0.0 & 0 & 5 & 0.0 & 0 \\
\cline{1-14}
\multirow[t]{3}{*}{25} & 0.25 & 5 & 0.0 & 1 & 5 & 0.0 & 0 & 5 & 0.0 & 0 & 5 & 0.0 & 0 \\
 & 0.50 & 5 & 0.0 & 24 & 5 & 0.0 & 26 & 5 & 0.0 & 37 & 5 & 0.0 & 33 \\
 & 0.75 & 5 & 0.0 & 1 & 5 & 0.0 & 0 & 5 & 0.0 & 0 & 5 & 0.0 & 0 \\
\cline{1-14}
\multirow[t]{3}{*}{30} & 0.25 & 5 & 0.0 & 56 & 5 & 0.0 & 41 & 5 & 0.0 & 64 & 5 & 0.0 & 112 \\
 & 0.50 & 0 & 15.4 & 901 & 0 & 18.7 & 901 & 0 & 15.2 & 901 & 0 & 14.6 & 900 \\
 & 0.75 & 4 & 1.0 & 555 & 5 & 0.0 & 406 & 4 & 1.1 & 546 & 5 & 0.0 & 499 \\
\cline{1-14}
\multirow[t]{3}{*}{35} & 0.25 & 0 & 17.2 & 901 & 0 & 16.2 & 900 & 0 & 14.9 & 900 & 0 & 15.1 & 900 \\
 & 0.50 & 0 & 33.8 & 901 & 0 & 30.4 & 900 & 0 & 27.8 & 900 & 0 & 28.1 & 900 \\
 & 0.75 & 0 & 12.6 & 901 & 0 & 14.2 & 900 & 0 & 13.9 & 900 & 0 & 14.2 & 900 \\
\cline{1-14}
\multirow[t]{3}{*}{40} & 0.25 & 0 & 58.7 & 901 & 0 & 33.4 & 900 & 0 & 39.5 & 900 & 0 & 34.1 & 900 \\
 & 0.50 & 0 & 66.3 & 900 & 0 & 42.0 & 900 & 0 & 37.5 & 900 & 0 & 36.9 & 900 \\
 & 0.75 & 0 & 45.3 & 901 & 0 & 29.8 & 900 & 0 & 29.5 & 900 & 0 & 20.9 & 900 \\
\cline{1-14}
 & Mean & 2.60 & 16.7 & 463 & 2.67 & 12.3 & 452 & 2.60 & 12.0 & 463 & 2.67 & 10.9 & 463 \\
\cline{1-14}
\bottomrule
\end{tabular}
}
\end{table}

\begin{figure}[!htbp]
     \centering
     \resizebox{0.98\textwidth}{!}{
     \begin{subfigure}{0.47\textwidth}
         \centering
         \includegraphics[width=\textwidth]{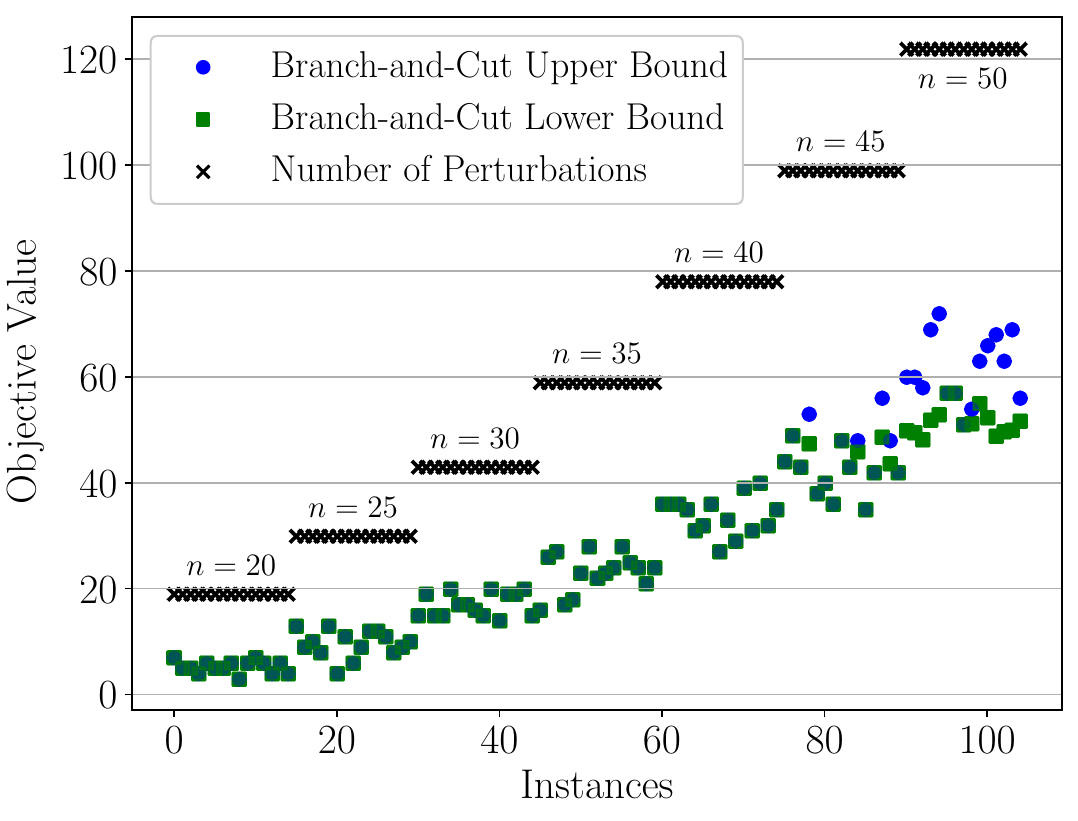} 
         \caption{Perturbation ratio 0.1.}
         \label{fig:edit_obj_comparison_01}
     \end{subfigure}
     \hspace{1em}
     \begin{subfigure}{0.47\textwidth}
         \centering
         \includegraphics[width=\textwidth]{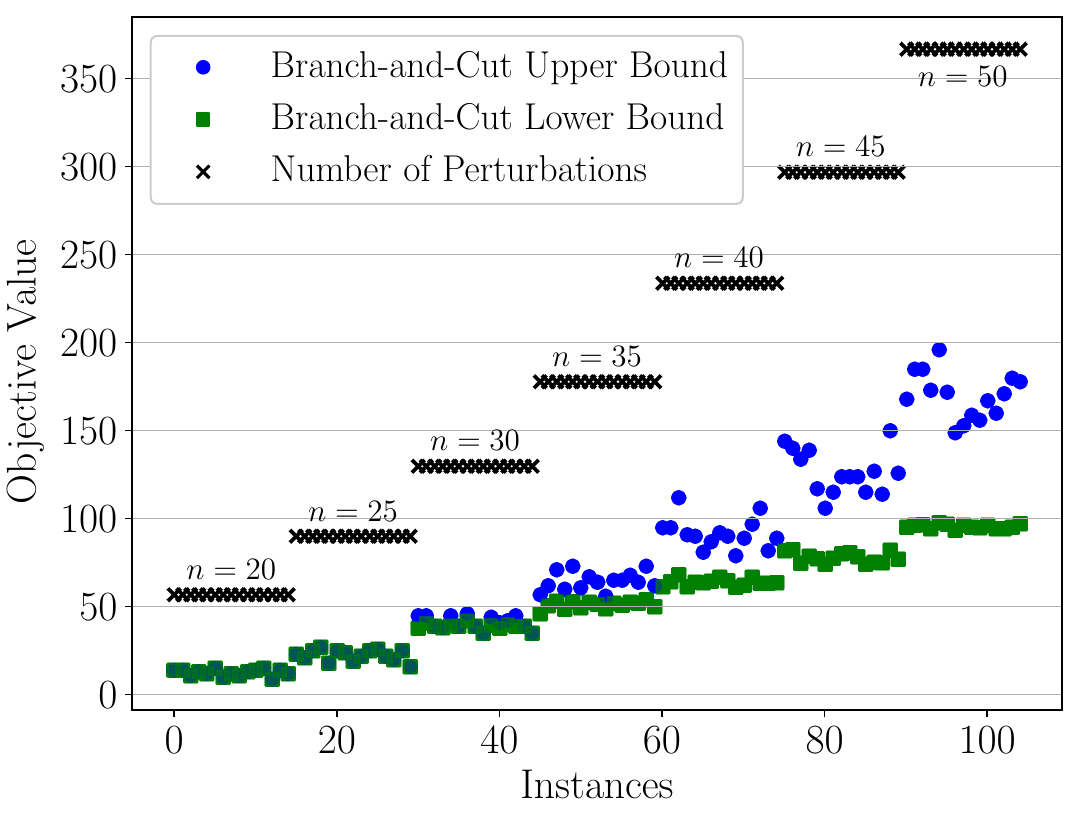} 
         \caption{Perturbation ratio 0.3.}
         \label{fig:edit_obj_comparison_03}
     \end{subfigure}
     }
    \caption{Upper bounds and lower bounds versus the number of perturbations on the perturbed perfect graphs for the minimum editing problem.}
     \label{fig:edit_obj_comparison}
 \end{figure}

Next, we evaluate the perturbed perfect graph instances. The detailed results are presented in Tables \ref{experiment_edit_strat_perturbed_01} and \ref{experiment_edit_strat_perturbed_03}.
Comparing Table \ref{experiment_edit_strat_perturbed_01} with Table~\ref{experiment_edit_strat_perturbed_03} reveals that the problem becomes more difficult to solve as the perturbation ratio increases. This is expected, as fewer perturbations yield fewer odd holes and odd antiholes, resulting in a graph that is closer to being perfect. We also observe that the results on perturbed perfect graphs with a perturbation ratio of 0.3 (Table~\ref{experiment_edit_strat_perturbed_03}) are similar to those of \erdosrenyi{} graphs (Table~\ref{experiment_edit_strat_er}) in terms of the number of instances solved to optimality and the optimality gaps for graphs of similar orders ($n \simeq 45$). This supports our intuition that perturbed perfect graph instances with perturbation ratio smaller than 0.1 are easier to solve.
Figures~\ref{fig:edit_obj_comparison_01} and \ref{fig:edit_obj_comparison_03} plot the initial number of perturbations alongside the best-known upper and lower bounds for these instances. While the original number of perturbations serves as a valid upper bound, the figures reveal that the optimal solutions require considerably fewer modifications. This indicates that the solver successfully identifies a much shorter sequence of edits to restore the graph to a perfect state than the sequence originally used to perturb it. 

\begin{table}[!htbp]
\centering
\setlength{\tabcolsep}{4pt}
\caption{Experimental results for the minimum perfect editing problem on perturbed perfect graphs.}
\label{experiment_edit_strat_perturbed_combined}

\begin{subtable}{\textwidth}
\centering
\caption{Perturbed perfect graphs with 0.1 perturbation ratio.}
\label{experiment_edit_strat_perturbed_01}
\resizebox{0.98\textwidth}{!}{
\begin{tabular}{cc|rrr|rrr|rrr|rrr|rrr}
\toprule
 &  & \multicolumn{12}{c}{Strategies} \\
 &  & \multicolumn{3}{c}{Base} & \multicolumn{3}{c}{H on C} & \multicolumn{3}{c}{H on R} & \multicolumn{3}{c}{H on C\&R} \\
$n$ & $d$ & Solved & Gap\% & Time(s) & Solved & Gap\% & Time(s) & Solved & Gap\% & Time(s) & Solved & Gap\% & Time(s) \\
\midrule
\multirow[t]{3}{*}{30} & 0.25 & 5 & 0.0 & 0 & 5 & 0.0 & 0 & 5 & 0.0 & 0 & 5 & 0.0 & 0 \\
 & 0.50 & 5 & 0.0 & 0 & 5 & 0.0 & 0 & 5 & 0.0 & 0 & 5 & 0.0 & 0 \\
 & 0.75 & 5 & 0.0 & 1 & 5 & 0.0 & 1 & 5 & 0.0 & 1 & 5 & 0.0 & 1 \\
\cline{1-14}
\multirow[t]{3}{*}{35} & 0.25 & 5 & 0.0 & 1 & 5 & 0.0 & 1 & 5 & 0.0 & 1 & 5 & 0.0 & 1 \\
 & 0.50 & 5 & 0.0 & 0 & 5 & 0.0 & 1 & 5 & 0.0 & 0 & 5 & 0.0 & 1 \\
 & 0.75 & 5 & 0.0 & 10 & 5 & 0.0 & 9 & 5 & 0.0 & 8 & 5 & 0.0 & 9 \\
\cline{1-14}
\multirow[t]{3}{*}{40} & 0.25 & 5 & 0.0 & 86 & 5 & 0.0 & 62 & 5 & 0.0 & 69 & 5 & 0.0 & 70 \\
 & 0.50 & 5 & 0.0 & 23 & 5 & 0.0 & 23 & 5 & 0.0 & 33 & 5 & 0.0 & 30 \\
 & 0.75 & 5 & 0.0 & 165 & 5 & 0.0 & 201 & 5 & 0.0 & 134 & 5 & 0.0 & 132 \\
\cline{1-14}
\multirow[t]{3}{*}{45} & 0.25 & 4 & 2.4 & 378 & 4 & 2.1 & 431 & 4 & 2.3 & 423 & 4 & 2.1 & 337 \\
 & 0.50 & 4 & 0.7 & 208 & 5 & 0.0 & 170 & 4 & 0.8 & 200 & 4 & 0.9 & 202 \\
 & 0.75 & 3 & 5.5 & 513 & 3 & 4.6 & 554 & 3 & 4.5 & 504 & 3 & 4.5 & 503 \\
\cline{1-14}
\multirow[t]{3}{*}{50} & 0.25 & 0 & 24.1 & 900 & 0 & 23.4 & 900 & 0 & 21.1 & 900 & 0 & 20.5 & 900 \\
 & 0.50 & 3 & 3.8 & 466 & 3 & 3.0 & 439 & 3 & 3.6 & 454 & 3 & 3.6 & 463 \\
 & 0.75 & 0 & 29.8 & 900 & 0 & 26.5 & 900 & 0 & 22.0 & 900 & 0 & 21.0 & 900 \\
\cline{1-14}
 & Mean & 3.93 & 4.4 & 243 & 4.00 & 4.0 & 246 & 3.93 & 3.6 & 242 & 3.93 & 3.5 & 237 \\
\cline{1-14}
\bottomrule
\end{tabular}
}
\vspace{0.8em}
\end{subtable}
\begin{subtable}{\textwidth}
\centering
\caption{Perturbed perfect graphs with 0.3 perturbation ratio.}
\label{experiment_edit_strat_perturbed_03}
\resizebox{0.98\textwidth}{!}{
\begin{tabular}{cc|rrr|rrr|rrr|rrr|rrr}
\toprule
 &  & \multicolumn{12}{c}{Strategies} \\
 &  & \multicolumn{3}{c}{Base} & \multicolumn{3}{c}{H on C} & \multicolumn{3}{c}{H on R} & \multicolumn{3}{c}{H on C\&R} \\
$n$ & $d$ & Solved & Gap\% & Time(s) & Solved & Gap\% & Time(s) & Solved & Gap\% & Time(s) & Solved & Gap\% & Time(s) \\
\midrule
\multirow[t]{3}{*}{25} & 0.25 & 5 & 0.0 & 2 & 5 & 0.0 & 4 & 5 & 0.0 & 3 & 5 & 0.0 & 2 \\
 & 0.50 & 5 & 0.0 & 2 & 5 & 0.0 & 0 & 5 & 0.0 & 2 & 5 & 0.0 & 1 \\
 & 0.75 & 5 & 0.0 & 1 & 5 & 0.0 & 1 & 5 & 0.0 & 2 & 5 & 0.0 & 2 \\
\cline{1-14}
\multirow[t]{3}{*}{30} & 0.25 & 1 & 9.2 & 825 & 1 & 8.9 & 851 & 2 & 7.2 & 800 & 2 & 7.6 & 803 \\
 & 0.50 & 3 & 3.9 & 444 & 3 & 3.1 & 458 & 3 & 3.6 & 467 & 3 & 3.7 & 455 \\
 & 0.75 & 2 & 5.9 & 698 & 2 & 6.7 & 637 & 2 & 5.8 & 603 & 2 & 5.0 & 603 \\
\cline{1-14}
\multirow[t]{3}{*}{35} & 0.25 & 0 & 23.7 & 900 & 0 & 22.6 & 900 & 0 & 20.5 & 900 & 0 & 21.8 & 900 \\
 & 0.50 & 0 & 19.8 & 900 & 0 & 19.3 & 900 & 0 & 18.4 & 900 & 0 & 18.4 & 900 \\
 & 0.75 & 0 & 27.8 & 900 & 0 & 23.7 & 900 & 0 & 22.2 & 900 & 0 & 21.7 & 900 \\
\cline{1-14}
\multirow[t]{3}{*}{40} & 0.25 & 0 & 65.0 & 900 & 0 & 37.0 & 900 & 0 & 34.1 & 900 & 0 & 33.6 & 900 \\
 & 0.50 & 0 & 31.8 & 900 & 0 & 27.4 & 900 & 0 & 25.2 & 900 & 0 & 25.1 & 900 \\
 & 0.75 & 0 & 44.2 & 900 & 0 & 35.6 & 900 & 0 & 30.1 & 900 & 0 & 30.5 & 900 \\
\cline{1-14}
\multirow[t]{3}{*}{45} & 0.25 & 0 & 80.3 & 902 & 0 & 46.8 & 900 & 0 & 40.9 & 900 & 0 & 41.1 & 900 \\
 & 0.50 & 0 & 83.9 & 900 & 0 & 37.7 & 900 & 0 & 33.7 & 900 & 0 & 33.8 & 900 \\
 & 0.75 & 0 & 80.7 & 900 & 0 & 44.1 & 900 & 0 & 38.6 & 900 & 0 & 38.9 & 900 \\
\cline{1-14}
\multirow[t]{3}{*}{50} & 0.25 & 0 & 80.6 & 901 & 0 & 57.3 & 903 & 0 & 47.6 & 900 & 0 & 46.9 & 900 \\
 & 0.50 & 0 & 84.1 & 901 & 0 & 43.1 & 900 & 0 & 40.2 & 900 & 0 & 39.4 & 900 \\
 & 0.75 & 0 & 81.1 & 901 & 0 & 53.5 & 902 & 0 & 44.2 & 901 & 0 & 44.1 & 900 \\
\cline{1-14}
 & Mean & 1.17 & 40.1 & 710 & 1.17 & 25.9 & 709 & 1.22 & 22.9 & 704 & 1.22 & 22.9 & 704 \\
\cline{1-14}
\bottomrule
\end{tabular}
}
\end{subtable}

\end{table}

The computational results for the $C_5$-free graph instances are provided in Table~\ref{experiment_edit_strat_c5}. The performance of the branch-and-cut algorithm and the relative effectiveness of the heuristic strategies on these instances align closely with the trends observed in the preceding experiments, further confirming the consistent behavior of our methods across different graph structures.
Our primary observation is that the branch-and-cut algorithm is able to solve instances of higher order ($n=75$) compared to \erdosrenyi{} and perturbed perfect graphs for the same optimality gaps and numbers of instances solved to optimality. This confirms our intuition that $C_5$-free graphs are easier for \textsc{MinPerfEdit}. Up to graph order 40, the algorithm is able to reach an optimal solution almost instantly. Over graph order 60, we observe that the algorithm is unable to conclude some instances due to memory exhaustion. Consequently, for orders exceeding 60, the `Solved' column reports the ratio of optimally solved instances to the total number of instances that successfully terminated without a memory error.

\begin{table}[!ht]
\centering
\setlength{\tabcolsep}{4pt}
\caption{Experimental results for the minimum perfect editing problem on $C_5$-free graphs using various strategy settings of \texttt{IterativeModificationHeuristic}.}
\label{experiment_edit_strat_c5}
\resizebox{1\textwidth}{!}{
\begin{tabular}{cc|rrr|rrr|rrr|rrr|rrr}
\toprule
 &  & \multicolumn{12}{c}{Strategies} \\
 &  & \multicolumn{3}{c}{Base} & \multicolumn{3}{c}{H on C} & \multicolumn{3}{c}{H on R} & \multicolumn{3}{c}{H on C\&R} \\
$n$ & $d$ & Solved & Gap\% & Time(s) & Solved & Gap\% & Time(s) & Solved & Gap\% & Time(s) & Solved & Gap\% & Time(s) \\
\midrule
\multirow[t]{3}{*}{40} & 0.25 & 5 & 0.0 & 0 & 5 & 0.0 & 0 & 5 & 0.0 & 0 & 5 & 0.0 & 0 \\
 & 0.50 & 5 & 0.0 & 0 & 5 & 0.0 & 0 & 5 & 0.0 & 0 & 5 & 0.0 & 0 \\
 & 0.75 & 5 & 0.0 & 3 & 5 & 0.0 & 3 & 5 & 0.0 & 3 & 5 & 0.0 & 3 \\
\cline{1-14}
\multirow[t]{3}{*}{45} & 0.25 & 4 & 18.3 & 188 & 4 & 3.7 & 185 & 4 & 3.4 & 185 & 4 & 3.4 & 185 \\
 & 0.50 & 5 & 0.0 & 0 & 5 & 0.0 & 0 & 5 & 0.0 & 0 & 5 & 0.0 & 0 \\
 & 0.75 & 5 & 0.0 & 7 & 5 & 0.0 & 31 & 5 & 0.0 & 8 & 5 & 0.0 & 31 \\
\cline{1-14}
\multirow[t]{3}{*}{50} & 0.25 & 5 & 0.0 & 14 & 5 & 0.0 & 13 & 5 & 0.0 & 12 & 5 & 0.0 & 12 \\
 & 0.50 & 5 & 0.0 & 0 & 5 & 0.0 & 0 & 5 & 0.0 & 0 & 5 & 0.0 & 0 \\
 & 0.75 & 5 & 0.0 & 72 & 5 & 0.0 & 55 & 5 & 0.0 & 36 & 5 & 0.0 & 36 \\
\cline{1-14}
\multirow[t]{3}{*}{55} & 0.25 & 5 & 0.0 & 58 & 5 & 0.0 & 60 & 5 & 0.0 & 33 & 5 & 0.0 & 35 \\
 & 0.50 & 5 & 0.0 & 0 & 5 & 0.0 & 0 & 5 & 0.0 & 0 & 5 & 0.0 & 0 \\
 & 0.75 & 5 & 0.0 & 143 & 5 & 0.0 & 147 & 5 & 0.0 & 142 & 5 & 0.0 & 147 \\
\cline{1-14}
\multirow[t]{3}{*}{60} & 0.25 & 5 & 0.0 & 59 & 5 & 0.0 & 40 & 5 & 0.0 & 59 & 5 & 0.0 & 41 \\
 & 0.50 & 5 & 0.0 & 10 & 5 & 0.0 & 7 & 5 & 0.0 & 5 & 5 & 0.0 & 4 \\
 & 0.75 & 4 & 4.9 & 216 & 4 & 4.3 & 218 & 4 & 2.6 & 213 & 4 & 4.3 & 219 \\
\cline{1-14}
\multirow[t]{3}{*}{65} & 0.25 & 2 / 4 & 48.4 & 527 & 1 / 3 & 19.4 & 627 & 3 / 4 & 7.1 & 373 & 2 / 3 & 8.5 & 423 \\
 & 0.50 & 5 / 5 & 0.0 & 2 & 5 / 5 & 0.0 & 3 & 5 / 5 & 0.0 & 2 & 5 / 5 & 0.0 & 2 \\
 & 0.75 & 5 / 5 & 0.0 & 68 & 5 / 5 & 0.0 & 69 & 5 / 5 & 0.0 & 68 & 5 / 5 & 0.0 & 69 \\
\cline{1-14}
\multirow[t]{3}{*}{70} & 0.25 & 3 / 3 & 0.0 & 58 & 3 / 3 & 0.0 & 58 & 3 / 3 & 0.0 & 58 & 3 / 3 & 0.0 & 58 \\
 & 0.50 & 5 / 5 & 0.0 & 2 & 5 / 5 & 0.0 & 2 & 5 / 5 & 0.0 & 2 & 5 / 5 & 0.0 & 2 \\
 & 0.75 & 2 / 3 & 15.7 & 429 & 2 / 3 & 4.2 & 412 & 2 / 3 & 3.2 & 432 & 2 / 3 & 3.0 & 413 \\
\cline{1-14}
\multirow[t]{2}{*}{75} & 0.50 & 5 / 5 & 0.0 & 20 & 5 / 5 & 0.0 & 12 & 5 / 5 & 0.0 & 10 & 5 / 5 & 0.0 & 12 \\
 & 0.75 & 3 / 3 & 0.0 & 283 & 3 / 3 & 0.0 & 291 & 3 / 3 & 0.0 & 288 & 3 / 3 & 0.0 & 295 \\
\cline{1-14}
 & Mean & 4.48 & 3.8 & 94 & 4.43 & 1.4 & 97 & 4.52 & 0.7 & 84 & 4.48 & 0.8 & 86 \\
\cline{1-14}
\bottomrule
\end{tabular}
}
\end{table}

\pagebreak

Finally, we present the experimental results on some of the well known DIMACS instances \citep{dimacs_challenges_archive} in Table~\ref{experiment_edit_dimacs}. Along with the instance name, we report the number of vertices ($n$), the number of edges ($m$), the number of odd holes (\#$OH$) and odd antiholes (\#$OAH$) present in the graph, the optimal objective value or the obtained lower and upper bounds and the optimality gap for the instance and the program runtime in seconds. To maintain the readability within the table, we restrict our results to the best known values and the strategy used to obtain these values. Unlike the experiments reported above, these results include graphs that have comparatively high orders solved to optimality with low runtime. The results demonstrate that the number of present odd holes and odd antiholes and their distribution in the graph have a greater impact on the practical difficulty of the problem than the graph's order and size. Likely due to the structural differences among these instances, these are not a single dominant strategy. However, applying the heuristic has a more pronounced effect on the instances that are harder to solve. In Section~\ref{sec:exp_imh}, we examine the results of the \texttt{IMH} on these DIMACS instances, providing further insight.

\begin{table}[H]
\centering
\setlength{\tabcolsep}{4pt}
\caption{Experimental results for the minimum perfect editing problem on some DIMACS instances.}
\label{experiment_edit_dimacs}
\resizebox{0.85\textwidth}{!}{
\begin{tabular}{lrrrrrrl}
\toprule
Instance & $n$ & $m$ & \#$OH$ & \#$OAH$ & Objective & Time(s) & Best Strategy \\
\midrule
anna & 138 & 493 & 202 & 0 & 15 & 0.8 & H on R \\
david & 87 & 406 & 953 & 0 & 16 & 1.0 & Base \\
fpsol2.i.2 & 451 & 8691 & 30016 & 0 & 28 & 43.0 & H on R \\
fpsol2.i.3 & 425 & 8688 & 30016 & 0 & 28 & 35.0 & Base \\
huck & 74 & 301 & 3 & 0 & 2 & 0.1 & Base \\
inithx.i.1 & 864 & 18707 & 16 & 0 & 1 & 239.7 & Base \\
jean & 80 & 254 & 67 & 0 & 6 & 0.1 & H on C\&R \\
miles1500 & 128 & 5198 & 72532 & 1811 & 24 & 3.4 & H on R \\
miles250 & 128 & 387 & 922 & 0 & 13 & 28.1 & Base \\
mulsol.i.2 & 188 & 3885 & 14336 & 0 & 25 & 5.4 & Base \\
mulsol.i.3 & 184 & 3916 & 14336 & 0 & 25 & 3.3 & H on R \\
mulsol.i.4 & 185 & 3946 & 14336 & 0 & 25 & 3.6 & Base \\
mulsol.i.5 & 186 & 3973 & 14336 & 0 & 25 & 2.2 & H on R \\
myciel3 & 11 & 20 & 31 & 0 & 4 & 0.0 & H on R \\
myciel4 & 23 & 71 & 646 & 0 & 16 & 0.2 & H on R \\
myciel5 & 47 & 236 & 17277 & 0 & 56 & 189.6 & Base \\
queen5\_5 & 25 & 160 & 600 & 136 & 31.48 -- 39 (19.29\%) & 903.0 & H on R \\
queen6\_6 & 36 & 290 & 10160 & 456 & 61.82 -- 91 (32.06\%) & 900.3 & H on C\&R \\
queen7\_7 & 49 & 476 & 120288 & 1056 & 98.4 -- 253 (61.11\%) & 924.1 & H on C\&R \\
queen8\_8 & 64 & 728 & 1685760 & 1960 & 145.6 -- 728 (80.0\%) & 912.3 & H on C\&R \\
zeroin.i.2 & 211 & 3541 & 102640 & 0 & 70 & 84.0 & Base \\
zeroin.i.3 & 206 & 3540 & 102640 & 0 & 70 & 80.1 & Base \\
\bottomrule
\end{tabular}
}
\end{table}

\subsection{The Minimum Perfect Completion Problem} \label{sec:completion}


In \textsc{MinPerfComp}, edge removal is strictly forbidden; the only permissible operation is the addition of edges between non-adjacent vertices. As previously noted, this problem is computationally equivalent to \textsc{MinPerfDel}, where only edge deletions are allowed. This equivalence is a direct consequence of the Weak Perfect Graph Theorem~\citep{weak_perfect}, which establishes that a graph is perfect if and only if its complement is perfect. Consequently, any instance of \textsc{MinPerfDel} can be solved by taking the complement of the input graph and solving \textsc{MinPerfComp}. The edges added to achieve perfectness in the complement correspond exactly to the edges that must be removed from the original graph. For this reason, we restrict our computational analysis to \textsc{MinPerfComp} and do not explicitly evaluate \textsc{MinPerfDel}.

To adapt our framework for \textsc{MinPerfComp}, we modify Model \texttt{IP\_Perfect} by defining decision variables exclusively for non-adjacent vertex pairs, as existing edges cannot be removed. The \texttt{IMH} is similarly customized to prevent the removal of original edges. With these modifications in place, we evaluate the branch-and-cut algorithm using the same heuristic application strategies explored for \textsc{MinPerfEdit} in Section~\ref{sec:edit}. Consistent with the conclusions drawn from our previous experiments, we set the \texttt{OHTP} parameter to \texttt{All}.

\begin{table}[!b]
\centering
\setlength{\tabcolsep}{4pt}
\caption{Experimental results for the minimum perfect completion problem on Erd\H{o}s--R\'enyi graphs using various strategy settings of \texttt{IterativeModificationHeuristic}.}
\label{experiment_comp_strat_er}
\resizebox{0.9\textwidth}{!}{
\begin{tabular}{cc|rrr|rrr|rrr|rrr|rrr}
\toprule
 &  & \multicolumn{12}{c}{Strategies} \\
 &  & \multicolumn{3}{c}{Base} & \multicolumn{3}{c}{H on C} & \multicolumn{3}{c}{H on R} & \multicolumn{3}{c}{H on C\&R} \\
$n$ & $d$ & Solved & Gap\% & Time(s) & Solved & Gap\% & Time(s) & Solved & Gap\% & Time(s) & Solved & Gap\% & Time(s) \\
\midrule
\multirow[t]{3}{*}{20} & 0.25 & 5 & 0.0 & 0 & 5 & 0.0 & 0 & 5 & 0.0 & 0 & 5 & 0.0 & 0 \\
 & 0.50 & 5 & 0.0 & 0 & 5 & 0.0 & 0 & 5 & 0.0 & 0 & 5 & 0.0 & 0 \\
 & 0.75 & 5 & 0.0 & 0 & 5 & 0.0 & 0 & 5 & 0.0 & 0 & 5 & 0.0 & 0 \\
\cline{1-14}
\multirow[t]{3}{*}{25} & 0.25 & 5 & 0.0 & 0 & 5 & 0.0 & 0 & 5 & 0.0 & 0 & 5 & 0.0 & 0 \\
 & 0.50 & 5 & 0.0 & 87 & 5 & 0.0 & 79 & 5 & 0.0 & 98 & 5 & 0.0 & 126 \\
 & 0.75 & 5 & 0.0 & 0 & 5 & 0.0 & 0 & 5 & 0.0 & 0 & 5 & 0.0 & 0 \\
\cline{1-14}
\multirow[t]{3}{*}{30} & 0.25 & 5 & 0.0 & 71 & 5 & 0.0 & 19 & 5 & 0.0 & 32 & 5 & 0.0 & 27 \\
 & 0.50 & 0 & 36.4 & 901 & 0 & 31.5 & 901 & 0 & 23.9 & 901 & 0 & 24.1 & 901 \\
 & 0.75 & 5 & 0.0 & 79 & 5 & 0.0 & 50 & 5 & 0.0 & 86 & 5 & 0.0 & 87 \\
\cline{1-14}
\multirow[t]{3}{*}{35} & 0.25 & 0 & 83.7 & 901 & 0 & 31.1 & 900 & 0 & 21.6 & 900 & 0 & 22.4 & 901 \\
 & 0.50 & 0 & 69.0 & 901 & 0 & 45.6 & 900 & 0 & 37.9 & 901 & 0 & 37.3 & 900 \\
 & 0.75 & 2 & 4.5 & 594 & 3 & 2.3 & 519 & 2 & 4.6 & 600 & 2 & 4.8 & 617 \\
\cline{1-14}
\multirow[t]{3}{*}{40} & 0.25 & 0 & 83.7 & 900 & 0 & 39.4 & 900 & 0 & 30.7 & 901 & 0 & 30.8 & 900 \\
 & 0.50 & 0 & 79.1 & 900 & 0 & 58.9 & 900 & 0 & 49.3 & 901 & 0 & 49.3 & 900 \\
 & 0.75 & 0 & 16.9 & 900 & 1 & 17.0 & 865 & 0 & 16.9 & 900 & 0 & 13.7 & 901 \\
\cline{1-14}
 & Mean & 2.80 & 24.9 & 416 & 2.93 & 15.1 & 402 & 2.80 & 12.3 & 415 & 2.80 & 12.2 & 417 \\
\cline{1-14}
\bottomrule
\end{tabular}
}
\end{table}

Table \ref{experiment_comp_strat_er} presents the results for \textsc{MinPerfComp} on \erdosrenyi{} graphs. The best strategy is the `H on C\&R' option, as in the previous experiments. The optimality gaps in dense graphs are significantly lower than in sparse graphs. This suggests that the completion problem is practically easier in dense graphs. We attribute this to the fact that our only available modification is edge addition, which increases the graph density. Denser graphs are closer to complete graphs. Moreover, there are fewer non-edges in dense graphs, thus fewer decision variables are defined in the corresponding integer program. We also observe that in both problems graphs of 0.5 density are harder compared to those with 0.25 and 0.75 densities. This can be explained by the higher number of odd holes and odd holes initially present in the \erdosrenyi{} graphs of density 0.5 and low graph order (see Figure~\ref{fig:E_X_Xbar_on_p_line_for_n}).

We also evaluated the performance of our branch-and-cut algorithm for \textsc{MinPerfComp} on the perturbed perfect graphs; see the computational results in Tables~\ref{experiment_comp_strat_perturbed_01} and \ref{experiment_comp_strat_perturbed_03}.
\begin{table}[H]
\centering
\setlength{\tabcolsep}{4pt}
\caption{Experimental results for the minimum perfect completion problem on perturbed perfect graphs.}
\label{experiment_comp_strat_perturbed_combined}
\begin{subtable}{\textwidth}
\centering
\caption{Perturbed perfect graphs with 0.1 perturbation ratio.}
\label{experiment_comp_strat_perturbed_01}
\resizebox{0.82\textwidth}{!}{
\begin{tabular}{cc|rrr|rrr|rrr|rrr|rrr}
\toprule
 &  & \multicolumn{12}{c}{Strategies} \\
 &  & \multicolumn{3}{c}{Base} & \multicolumn{3}{c}{H on C} & \multicolumn{3}{c}{H on R} & \multicolumn{3}{c}{H on C\&R} \\
$n$ & $d$ & Solved & Gap\% & Time(s) & Solved & Gap\% & Time(s) & Solved & Gap\% & Time(s) & Solved & Gap\% & Time(s) \\
\midrule
\multirow[t]{3}{*}{30} & 0.25 & 5 & 0.0 & 5 & 5 & 0.0 & 6 & 5 & 0.0 & 10 & 5 & 0.0 & 9 \\
 & 0.50 & 5 & 0.0 & 0 & 5 & 0.0 & 0 & 5 & 0.0 & 0 & 5 & 0.0 & 1 \\
 & 0.75 & 5 & 0.0 & 0 & 5 & 0.0 & 0 & 5 & 0.0 & 0 & 5 & 0.0 & 0 \\
\cline{1-14}
\multirow[t]{3}{*}{35} & 0.25 & 4 & 3.8 & 302 & 4 & 2.6 & 373 & 4 & 3.0 & 341 & 4 & 1.9 & 377 \\
 & 0.50 & 5 & 0.0 & 2 & 5 & 0.0 & 1 & 5 & 0.0 & 2 & 5 & 0.0 & 2 \\
 & 0.75 & 5 & 0.0 & 5 & 5 & 0.0 & 5 & 5 & 0.0 & 9 & 5 & 0.0 & 9 \\
\cline{1-14}
\multirow[t]{3}{*}{40} & 0.25 & 0 & 79.1 & 901 & 0 & 32.7 & 900 & 0 & 25.1 & 900 & 0 & 24.2 & 901 \\
 & 0.50 & 3 & 2.5 & 453 & 4 & 1.5 & 453 & 2 & 4.5 & 548 & 3 & 3.0 & 537 \\
 & 0.75 & 4 & 0.6 & 259 & 4 & 0.9 & 334 & 4 & 1.6 & 302 & 4 & 1.5 & 289 \\
\cline{1-14}
\multirow[t]{3}{*}{45} & 0.25 & 0 & 59.9 & 900 & 0 & 30.9 & 900 & 0 & 23.9 & 901 & 0 & 24.2 & 902 \\
 & 0.50 & 2 & 22.2 & 638 & 2 & 9.6 & 619 & 2 & 9.9 & 667 & 2 & 9.0 & 707 \\
 & 0.75 & 2 & 5.6 & 586 & 2 & 5.1 & 573 & 2 & 6.1 & 596 & 2 & 5.5 & 646 \\
\cline{1-14}
\multirow[t]{3}{*}{50} & 0.25 & 0 & 87.4 & 900 & 0 & 48.6 & 900 & 0 & 38.3 & 901 & 0 & 38.2 & 901 \\
 & 0.50 & 0 & 61.0 & 900 & 0 & 30.2 & 900 & 0 & 25.7 & 902 & 0 & 24.5 & 902 \\
 & 0.75 & 0 & 29.2 & 900 & 0 & 26.7 & 901 & 0 & 22.3 & 901 & 0 & 21.9 & 901 \\
\cline{1-14}
 & Mean & 2.67 & 23.4 & 450 & 2.73 & 12.6 & 458 & 2.60 & 10.7 & 465 & 2.67 & 10.3 & 472 \\
\cline{1-14}
\bottomrule
\end{tabular}
}
\end{subtable}

\vspace{0.8em}

\begin{subtable}{\textwidth}
\centering
\caption{Perturbed perfect graphs with 0.3 perturbation ratio.}
\label{experiment_comp_strat_perturbed_03}
\resizebox{0.82\textwidth}{!}{
\begin{tabular}{cc|rrr|rrr|rrr|rrr|rrr}
\toprule
 &  & \multicolumn{12}{c}{Strategies} \\
 &  & \multicolumn{3}{c}{Base} & \multicolumn{3}{c}{H on C} & \multicolumn{3}{c}{H on R} & \multicolumn{3}{c}{H on C\&R} \\
$n$ & $d$ & Solved & Gap\% & Time(s) & Solved & Gap\% & Time(s) & Solved & Gap\% & Time(s) & Solved & Gap\% & Time(s) \\
\midrule
\multirow[t]{3}{*}{25} & 0.25 & 5 & 0.0 & 12 & 5 & 0.0 & 10 & 5 & 0.0 & 20 & 5 & 0.0 & 16 \\
 & 0.50 & 5 & 0.0 & 7 & 5 & 0.0 & 7 & 5 & 0.0 & 9 & 5 & 0.0 & 9 \\
 & 0.75 & 5 & 0.0 & 6 & 5 & 0.0 & 5 & 5 & 0.0 & 10 & 5 & 0.0 & 11 \\
\cline{1-14}
\multirow[t]{3}{*}{30} & 0.25 & 0 & 19.0 & 900 & 0 & 16.2 & 900 & 0 & 13.2 & 901 & 0 & 12.7 & 901 \\
 & 0.50 & 2 & 9.1 & 630 & 2 & 11.4 & 634 & 2 & 9.2 & 661 & 2 & 9.1 & 690 \\
 & 0.75 & 0 & 14.2 & 900 & 1 & 11.7 & 803 & 0 & 11.8 & 901 & 0 & 13.3 & 901 \\
\cline{1-14}
\multirow[t]{3}{*}{35} & 0.25 & 0 & 59.5 & 900 & 0 & 36.0 & 900 & 0 & 30.5 & 900 & 0 & 30.1 & 901 \\
 & 0.50 & 0 & 34.1 & 900 & 0 & 31.4 & 900 & 0 & 25.1 & 900 & 0 & 24.4 & 900 \\
 & 0.75 & 0 & 53.3 & 900 & 0 & 37.3 & 900 & 0 & 32.0 & 900 & 0 & 32.6 & 900 \\
\cline{1-14}
\multirow[t]{3}{*}{40} & 0.25 & 0 & 77.6 & 900 & 0 & 50.1 & 900 & 0 & 43.4 & 900 & 0 & 42.4 & 900 \\
 & 0.50 & 0 & 68.8 & 901 & 0 & 41.0 & 900 & 0 & 34.5 & 900 & 0 & 34.1 & 901 \\
 & 0.75 & 0 & 74.0 & 900 & 0 & 44.6 & 900 & 0 & 41.7 & 900 & 0 & 40.0 & 900 \\
\cline{1-14}
\multirow[t]{3}{*}{45} & 0.25 & 0 & 79.5 & 900 & 0 & 52.0 & 900 & 0 & 45.8 & 901 & 0 & 45.2 & 903 \\
 & 0.50 & 0 & 78.8 & 900 & 0 & 53.9 & 900 & 0 & 46.7 & 901 & 0 & 46.3 & 901 \\
 & 0.75 & 0 & 78.8 & 900 & 0 & 55.5 & 900 & 0 & 50.4 & 900 & 0 & 49.4 & 900 \\
\cline{1-14}
\multirow[t]{3}{*}{50} & 0.25 & 0 & 79.5 & 901 & 0 & 63.0 & 902 & 0 & 51.5 & 909 & 0 & 52.0 & 906 \\
 & 0.50 & 0 & 79.4 & 901 & 0 & 66.9 & 900 & 0 & 51.8 & 907 & 0 & 52.8 & 904 \\
 & 0.75 & 0 & 78.9 & 901 & 0 & 71.3 & 901 & 0 & 55.7 & 902 & 0 & 55.1 & 903 \\
\cline{1-14}
 & Mean & 0.94 & 49.1 & 737 & 1.00 & 35.7 & 731 & 0.94 & 30.2 & 740 & 0.94 & 30.0 & 742 \\
\cline{1-14}
\bottomrule
\end{tabular}
}
\end{subtable}
\end{table}

Consistent with our earlier findings, the algorithmic behavior and the relative success of the heuristic strategies on perturbed perfect graphs align closely with the trends observed for the editing problem. Specifically, the instances once again become consistently harder to solve as the perturbation ratio increases. Moreover, a direct comparison of the solved instance counts against those of the editing problem (Tables \ref{experiment_edit_strat_perturbed_01} and \ref{experiment_edit_strat_perturbed_03}) reveals that the completion problem is computationally more demanding on these graphs.  Furthermore, the problem becomes progressively easier to solve as graph density increases.

\begin{table}[!b]
\centering
\setlength{\tabcolsep}{4pt}
\caption{Experimental results for the minimum perfect completion problem on $C_5$-free graphs using various strategy settings of \texttt{IterativeModificationHeuristic}.}
\label{experiment_comp_strat_c5}
\resizebox{0.9\textwidth}{!}{
\begin{tabular}{cc|rrr|rrr|rrr|rrr|rrr}
\toprule
 &  & \multicolumn{12}{c}{Strategies} \\
 &  & \multicolumn{3}{c}{Base} & \multicolumn{3}{c}{H on C} & \multicolumn{3}{c}{H on R} & \multicolumn{3}{c}{H on C\&R} \\
$n$ & $d$ & Solved & Gap\% & Time(s) & Solved & Gap\% & Time(s) & Solved & Gap\% & Time(s) & Solved & Gap\% & Time(s) \\
\midrule
\multirow[t]{3}{*}{25} & 0.25 & 5 & 0.0 & 0 & 5 & 0.0 & 0 & 5 & 0.0 & 0 & 5 & 0.0 & 0 \\
 & 0.50 & 5 & 0.0 & 0 & 5 & 0.0 & 0 & 5 & 0.0 & 0 & 5 & 0.0 & 0 \\
 & 0.75 & 5 & 0.0 & 0 & 5 & 0.0 & 0 & 5 & 0.0 & 0 & 5 & 0.0 & 0 \\
\cline{1-14}
\multirow[t]{3}{*}{30} & 0.25 & 4 & 3.8 & 253 & 5 & 0.0 & 104 & 4 & 3.7 & 297 & 5 & 0.0 & 236 \\
 & 0.50 & 5 & 0.0 & 0 & 5 & 0.0 & 0 & 5 & 0.0 & 1 & 5 & 0.0 & 0 \\
 & 0.75 & 5 & 0.0 & 0 & 5 & 0.0 & 0 & 5 & 0.0 & 0 & 5 & 0.0 & 0 \\
\cline{1-14}
\multirow[t]{3}{*}{35} & 0.25 & 3 & 8.7 & 403 & 3 & 9.1 & 417 & 4 & 5.3 & 429 & 4 & 1.5 & 388 \\
 & 0.50 & 5 & 0.0 & 0 & 5 & 0.0 & 0 & 5 & 0.0 & 0 & 5 & 0.0 & 0 \\
 & 0.75 & 5 & 0.0 & 0 & 5 & 0.0 & 0 & 5 & 0.0 & 0 & 5 & 0.0 & 0 \\
\cline{1-14}
\multirow[t]{3}{*}{40} & 0.25 & 4 & 7.6 & 235 & 4 & 1.9 & 271 & 4 & 6.1 & 221 & 4 & 4.2 & 221 \\
 & 0.50 & 5 & 0.0 & 0 & 5 & 0.0 & 0 & 5 & 0.0 & 0 & 5 & 0.0 & 0 \\
 & 0.75 & 5 & 0.0 & 0 & 5 & 0.0 & 0 & 5 & 0.0 & 0 & 5 & 0.0 & 0 \\
\cline{1-14}
\multirow[t]{3}{*}{45} & 0.25 & 3 & 38.3 & 388 & 3 & 22.2 & 408 & 3 & 17.9 & 390 & 3 & 18.3 & 408 \\
 & 0.50 & 5 & 0.0 & 82 & 5 & 0.0 & 85 & 5 & 0.0 & 53 & 5 & 0.0 & 76 \\
 & 0.75 & 5 & 0.0 & 1 & 5 & 0.0 & 1 & 5 & 0.0 & 1 & 5 & 0.0 & 1 \\
\cline{1-14}
\multirow[t]{3}{*}{50} & 0.25 & 0 & 96.6 & 937 & 0 & 39.0 & 963 & 0 & 52.8 & 949 & 0 & 35.6 & 910 \\
 & 0.50 & 4 & 12.5 & 187 & 4 & 4.6 & 193 & 4 & 5.8 & 203 & 4 & 8.2 & 220 \\
 & 0.75 & 5 & 0.0 & 2 & 5 & 0.0 & 2 & 5 & 0.0 & 3 & 5 & 0.0 & 3 \\
\cline{1-14}
\multirow[t]{3}{*}{55} & 0.25 & 0 & 82.8 & 908 & 0 & 43.2 & 921 & 1 & 38.5 & 746 & 1 & 33.9 & 745 \\
 & 0.50 & 4 & 19.4 & 355 & 4 & 13.3 & 249 & 4 & 11.5 & 318 & 4 & 11.2 & 247 \\
 & 0.75 & 5 & 0.0 & 7 & 5 & 0.0 & 7 & 5 & 0.0 & 6 & 5 & 0.0 & 6 \\
\cline{1-14}
\multirow[t]{3}{*}{60} & 0.25 & 1 / 4 & 73.6 & 921 & 3 / 4 & 24.4 & 405 & 1 / 4 & 58.6 & 780 & 3 / 4 & 24.4 & 404 \\
 & 0.50 & 2 / 5 & 47.1 & 577 & 2 / 5 & 28.4 & 551 & 2 / 5 & 25.2 & 557 & 2 / 5 & 23.5 & 555 \\
 & 0.75 & 5 / 5 & 0.0 & 9 & 5 / 5 & 0.0 & 9 & 5 / 5 & 0.0 & 9 & 5 / 5 & 0.0 & 10 \\
\cline{1-14}
\multirow[t]{3}{*}{65} & 0.25 & 0 / 2 & 97.8 & 1090 & 0 / 2 & 27.9 & 951 & 0 / 2 & 58.8 & 1080 & 0 / 2 & 27.9 & 950 \\
 & 0.50 & 2 / 5 & 58.6 & 820 & 2 / 5 & 39.7 & 671 & 2 / 5 & 35.9 & 785 & 2 / 5 & 36.9 & 665 \\
 & 0.75 & 5 / 5 & 0.0 & 12 & 5 / 5 & 0.0 & 12 & 5 / 5 & 0.0 & 13 & 5 / 5 & 0.0 & 13 \\
\cline{1-14}
\multirow[t]{3}{*}{70} & 0.25 & 1 / 1 & 0.0 & 126 & 1 / 1 & 0.0 & 461 & 1 / 1 & 0.0 & 126 & 1 / 1 & 0.0 & 461 \\
 & 0.50 & 2 / 5 & 58.7 & 542 & 2 / 5 & 33.0 & 544 & 2 / 5 & 33.9 & 543 & 2 / 5 & 30.9 & 548 \\
 & 0.75 & 3 / 3 & 0.0 & 44 & 3 / 3 & 0.0 & 46 & 3 / 3 & 0.0 & 46 & 3 / 3 & 0.0 & 48 \\
\cline{1-14}
\multirow[t]{2}{*}{75} & 0.50 & 2 / 5 & 58.8 & 547 & 3 / 5 & 27.0 & 569 & 2 / 5 & 24.1 & 610 & 3 / 5 & 18.2 & 629 \\
 & 0.75 & 5 / 5 & 0.0 & 50 & 5 / 5 & 0.0 & 52 & 5 / 5 & 0.0 & 54 & 5 / 5 & 0.0 & 57 \\
\cline{1-14}
 & Mean & 3.59 & 20.8 & 266 & 3.72 & 9.8 & 247 & 3.66 & 11.8 & 257 & 3.78 & 8.6 & 244 \\
\cline{1-14}
\bottomrule
\end{tabular}
}
\end{table}

Table \ref{experiment_comp_strat_c5} displays the results for \textsc{MinPerfComp} on $C_5$-free graphs. Comparing these with the \textsc{MinPerfEdit} experiments (Table~\ref{experiment_edit_strat_c5}) reveals that the minimum editing problem is computationally less demanding than the minimum completion problem on $C_5$-free graphs, similar to what we observed with perturbed perfect graphs. For \textsc{MinPerfComp}, we see a similar asymmetry between graphs of density 0.25 and 0.75; namely, the performance is better for graphs of 0.75 density. However, instances with a 0.5 density are easier to solve than those with a 0.25 density on $C_5$-free graphs, unlike in the editing problem. As in \textsc{MinPerfEdit} experiments, certain higher-order instances fail to conclude due to memory exhaustion.

Table~\ref{experiment_comp_strat_dimacs} demonstrates the results for  \textsc{MinPerfComp} on the DIMACS instances. Naturally, we see that the optimal objective values are higher than or equal to those in the editing problem (See Table~\ref{experiment_edit_dimacs}). Since the feasible solution set of a completion problem instance is a subset of the feasible solution set of the editing problem for the same graph, the optimal objective value of the editing problem is a lower bound for the minimum completion problem. Similar to the editing problem, there is no dominating strategy for DIMACS instances. 

\begin{table}[H]
\centering
\setlength{\tabcolsep}{4pt}
\caption{Experimental results for the minimum perfect completion problem on several DIMACS instances.}
\label{experiment_comp_strat_dimacs}
\resizebox{0.85\textwidth}{!}{
\begin{tabular}{lrrrrrrl}
\toprule
Instance & $n$ & $m$ & \#$OH$ & \#$OAH$ & Objective & Time(s) & Best Strategy \\
\midrule
anna & 138 & 493 & 202 & 0 & 20 & 0.5 & Base \\
david & 87 & 406 & 953 & 0 & 31 & 28.5 & Base \\
fpsol2.i.2 & 451 & 8691 & 30016 & 0 & 40 & 12.5 & Base \\
fpsol2.i.3 & 425 & 8688 & 30016 & 0 & 40 & 11.4 & H on R \\
huck & 74 & 301 & 3 & 0 & 2 & 0.1 & H on C\&R \\
inithx.i.1 & 864 & 18707 & 16 & 0 & 1 & 158.5 & H on R \\
jean & 80 & 254 & 67 & 0 & 6 & 0.1 & H on C\&R \\
miles1500 & 128 & 5198 & 72532 & 1811 & 24 & 2.2 & H on R \\
miles250 & 128 & 387 & 922 & 0 & 23 & 64.3 & Base \\
mulsol.i.2 & 188 & 3885 & 14336 & 0 & 35 & 1.0 & H on C\&R \\
mulsol.i.3 & 184 & 3916 & 14336 & 0 & 35 & 1.0 & H on C\&R \\
mulsol.i.4 & 185 & 3946 & 14336 & 0 & 35 & 1.0 & H on C\&R \\
mulsol.i.5 & 186 & 3973 & 14336 & 0 & 35 & 1.0 & H on C\&R \\
myciel3 & 11 & 20 & 31 & 0 & 6 & 0.1 & H on C\&R \\
myciel4 & 23 & 71 & 646 & 0 & 27 & 0.4 & H on C \\
myciel5 & 47 & 236 & 17277 & 0 & 103.04 -- 118 (12.68\%) & 900.7 & H on C\&R \\
queen5\_5 & 25 & 160 & 600 & 136 & 40 & 630.9 & H on C\&R \\
queen6\_6 & 36 & 290 & 10160 & 456 & 73.93 -- 166 (55.46\%) & 900.7 & H on C\&R \\
queen7\_7 & 49 & 476 & 120288 & 1056 & 142.4 -- 405 (64.84\%) & 927.3 & H on R \\
queen8\_8 & 64 & 728 & 1685760 & 1960 & 257.6 -- 865 (70.22\%) & 911.1 & H on C\&R \\
zeroin.i.2 & 211 & 3541 & 102640 & 0 & 104 & 3.4 & H on C\&R \\
zeroin.i.3 & 206 & 3540 & 102640 & 0 & 104 & 5.0 & H on R \\
\bottomrule
\end{tabular}
}
\end{table}

\subsection{The Perfect Sandwich Problem} \label{sec:sandwich}

Unlike the editing and completion problems, \textsc{PerfSand} is fundamentally a feasibility problem. Depending on the arrangement of mandatory and optional edges, a perfect graph satisfying the sandwich condition may not exist. Consequently, our objective is simply to determine if any perfect graph exists within this valid edge space, rather than minimizing the number of edge modifications. To align Model \texttt{IP\_Perfect} and our branch-and-cut algorithm with this objective, we define decision variables exclusively for the optional edges, as the statuses of all mandatory and strictly forbidden edges are fixed. Furthermore, because optimality is irrelevant, we configure the CPLEX solver to terminate immediately upon discovering the first feasible integer solution. Conversely, if no such graph exists within the bounds, the solver will exhaust the search space and mathematically prove the integer program to be infeasible.

Because the statuses of mandatory edges are fixed, any induced odd hole present in the input graph must be resolved by adding edges. If none of the missing edges for such a hole are included in the optional edge set, it is impossible to eliminate it, rendering the instance inherently infeasible. To exploit this structural limitation, we implemented a preprocessing mechanism called \texttt{Precheck}. This routine evaluates the available optional edges while generating the initial constraints for the odd holes and odd antiholes in the input graph. If \texttt{Precheck} identifies an odd hole or odd antihole for which no missing edges are optional, it immediately flags the instance as infeasible and terminates the algorithm. While highly efficient, this process is only a partial filter and does not capture all infeasible instances.

Because the existence of a perfect graph is not guaranteed in \textsc{PerfSand}, we do not apply the \texttt{IMH}. However, because this is purely a feasibility problem where the solver terminates immediately upon finding a valid solution, it is significantly less computationally demanding than the optimization variants. Consequently, we were able to evaluate \textsc{PerfSand} on graphs of considerably higher order. In addition to the \erdosrenyi{} graphs used in the editing and completion experiments, we generated larger instances with orders ranging from 50 to 130 in increments of 10. For each instance, optional edges are generated uniformly at random from the set of existing non-edges. In this context, we define \textit{optional edge density} as the proportion of an input graph's non-edges that are selected to become optional. Our preliminary experiments indicated that instances with an optional edge density of 0.5 or lower are almost universally infeasible. Therefore, we restricted our computational evaluation to optional edge densities of 0.70, 0.80, and 0.90.

In Table \ref{experiment_sandwich_70_80_90}, the `Time(s) in Precheck' column and `Time(s)' column represent the average runtimes of the program in the \texttt{Precheck} mechanism and total runtime of the program in seconds, respectively. The remaining columns indicate the final solver statuses out of the five instances tested: the number of feasible instances (`F'), instances proven infeasible by \texttt{Precheck} (`I$_1$'), instances proven infeasible by the branch-and-cut algorithm (`I$_2$'), and instances with an unknown status due to reaching the 900 seconds time limit (`U').

\begin{table}[!t]
\centering
\setlength{\tabcolsep}{4pt}
\caption{Experimental results for the perfect sandwich problem.}
\label{experiment_sandwich_70_80_90}
\resizebox{1\textwidth}{!}{
\begin{tabular}{cc|rrllll|rrllll|rrllll}
\toprule
 &  & \multicolumn{18}{c}{Optional Edge Density} \\
 &  & \multicolumn{6}{c}{0.70} & \multicolumn{6}{c}{0.80} & \multicolumn{6}{c}{0.90} \\
 $n$ & $d$ & \makecell{Time(s) in \\ Precheck} & Time(s) & F & I$_1$ & I$_2$ & U & \makecell{Time(s) in \\ Precheck} & Time(s) & F & I$_1$ & I$_2$ & U & \makecell{Time(s) in \\ Precheck} & Time(s) & F & I$_1$ & I$_2$ & U \\
\midrule
\multirow[t]{3}{*}{20} & 0.25 & 0.00 & 0.16 & 4 & 0 & 1 & 0 & 0.00 & 0.24 & 5 & 0 & 0 & 0 & 0.00 & 0.10 & 5 & 0 & 0 & 0 \\
 & 0.50 & 0.01 & 0.17 & 4 & 1 & 0 & 0 & 0.01 & 0.26 & 5 & 0 & 0 & 0 & 0.00 & 0.01 & 5 & 0 & 0 & 0 \\
 & 0.75 & 0.00 & 0.07 & 4 & 0 & 1 & 0 & 0.00 & 0.02 & 5 & 0 & 0 & 0 & 0.00 & 0.01 & 5 & 0 & 0 & 0 \\
\cline{1-20}
\multirow[t]{3}{*}{25} & 0.25 & 0.01 & 0.43 & 0 & 1 & 4 & 0 & 0.01 & 0.27 & 4 & 1 & 0 & 0 & 0.01 & 0.19 & 5 & 0 & 0 & 0 \\
 & 0.50 & 0.01 & 0.17 & 0 & 3 & 2 & 0 & 0.01 & 0.29 & 4 & 1 & 0 & 0 & 0.01 & 0.14 & 5 & 0 & 0 & 0 \\
 & 0.75 & 0.00 & 0.06 & 3 & 2 & 0 & 0 & 0.00 & 0.14 & 5 & 0 & 0 & 0 & 0.00 & 0.02 & 5 & 0 & 0 & 0 \\
\cline{1-20}
\multirow[t]{3}{*}{30} & 0.25 & 0.01 & 1.15 & 0 & 2 & 3 & 0 & 0.01 & 13.86 & 5 & 0 & 0 & 0 & 0.01 & 0.32 & 5 & 0 & 0 & 0 \\
 & 0.50 & 0.00 & 0.12 & 0 & 4 & 1 & 0 & 0.02 & 3.82 & 0 & 1 & 4 & 0 & 0.02 & 0.18 & 5 & 0 & 0 & 0 \\
 & 0.75 & 0.00 & 0.01 & 0 & 5 & 0 & 0 & 0.01 & 0.17 & 5 & 0 & 0 & 0 & 0.01 & 0.03 & 5 & 0 & 0 & 0 \\
\cline{1-20}
\multirow[t]{3}{*}{35} & 0.25 & 0.02 & 0.13 & 0 & 4 & 1 & 0 & 0.04 & 14.92 & 0 & 2 & 3 & 0 & 0.04 & 1.56 & 5 & 0 & 0 & 0 \\
 & 0.50 & 0.01 & 0.01 & 0 & 5 & 0 & 0 & 0.02 & 0.23 & 0 & 4 & 1 & 0 & 0.03 & 0.81 & 4 & 0 & 1 & 0 \\
 & 0.75 & 0.01 & 0.08 & 0 & 4 & 1 & 0 & 0.02 & 0.09 & 2 & 3 & 0 & 0 & 0.04 & 0.22 & 5 & 0 & 0 & 0 \\
\cline{1-20}
\multirow[t]{3}{*}{40} & 0.25 & 0.03 & 0.29 & 0 & 4 & 1 & 0 & 0.05 & 63.30 & 0 & 4 & 1 & 0 & 0.08 & 365.50 & 3 & 0 & 0 & 2 \\
 & 0.50 & 0.01 & 0.02 & 0 & 5 & 0 & 0 & 0.03 & 6.21 & 0 & 4 & 1 & 0 & 0.07 & 5.82 & 5 & 0 & 0 & 0 \\
 & 0.75 & 0.00 & 0.01 & 0 & 5 & 0 & 0 & 0.04 & 0.18 & 2 & 3 & 0 & 0 & 0.11 & 0.15 & 5 & 0 & 0 & 0 \\
\cline{1-20}
\multirow[t]{3}{*}{50} & 0.25 & 0.06 & 0.07 & 0 & 5 & 0 & 0 & 0.42 & 61.38 & 0 & 4 & 1 & 0 & 0.47 & 903.28 & 0 & 0 & 0 & 5 \\
 & 0.50 & 0.01 & 0.01 & 0 & 5 & 0 & 0 & 0.05 & 0.06 & 0 & 5 & 0 & 0 & 0.14 & 386.93 & 1 & 2 & 0 & 2 \\
 & 0.75 & 0.00 & 0.00 & 0 & 5 & 0 & 0 & 0.03 & 0.03 & 0 & 5 & 0 & 0 & 0.48 & 35.80 & 5 & 0 & 0 & 0 \\
\cline{1-20}
\multirow[t]{3}{*}{60} & 0.25 & 0.38 & 0.39 & 0 & 5 & 0 & 0 & 1.96 & 29.42 & 0 & 4 & 1 & 0 & --- & --- & --- & --- & --- & --- \\
 & 0.50 & 0.00 & 0.00 & 0 & 5 & 0 & 0 & 0.03 & 0.04 & 0 & 5 & 0 & 0 & 0.22 & 391.93 & 0 & 3 & 0 & 2 \\
 & 0.75 & 0.00 & 0.00 & 0 & 5 & 0 & 0 & 0.02 & 0.02 & 0 & 5 & 0 & 0 & 2.56 & 543.78 & 1 & 1 & 0 & 3 \\
\cline{1-20}
\multirow[t]{3}{*}{70} & 0.25 & 0.30 & 0.30 & 0 & 5 & 0 & 0 & 3.02 & 3.04 & 0 & 5 & 0 & 0 & 10.68 & 10.73 & 0 & 2 & 0 & 0 \\
 & 0.50 & 0.01 & 0.02 & 0 & 5 & 0 & 0 & 0.05 & 0.06 & 0 & 5 & 0 & 0 & 0.24 & 0.26 & 0 & 4 & 0 & 0 \\
 & 0.75 & 0.00 & 0.00 & 0 & 5 & 0 & 0 & 0.02 & 0.03 & 0 & 5 & 0 & 0 & --- & --- & --- & --- & --- & --- \\
\cline{1-20}
\multirow[t]{3}{*}{80} & 0.25 & 0.72 & 0.73 & 0 & 5 & 0 & 0 & 5.63 & 5.65 & 0 & 5 & 0 & 0 & 77.66 & 77.70 & 0 & 1 & 0 & 0 \\
 & 0.50 & 0.03 & 0.04 & 0 & 5 & 0 & 0 & 0.03 & 0.03 & 0 & 5 & 0 & 0 & 0.81 & 0.84 & 0 & 3 & 0 & 0 \\
 & 0.75 & 0.01 & 0.02 & 0 & 5 & 0 & 0 & 0.00 & 0.01 & 0 & 5 & 0 & 0 & --- & --- & --- & --- & --- & --- \\
\cline{1-20}
\multirow[t]{3}{*}{90} & 0.25 & 3.70 & 3.72 & 0 & 5 & 0 & 0 & 27.99 & 28.01 & 0 & 5 & 0 & 0 & 149.32 & 149.35 & 0 & 1 & 0 & 0 \\
 & 0.50 & 0.04 & 0.05 & 0 & 5 & 0 & 0 & 0.06 & 0.07 & 0 & 5 & 0 & 0 & 1.24 & 1.26 & 0 & 5 & 0 & 0 \\
 & 0.75 & 0.01 & 0.01 & 0 & 5 & 0 & 0 & 0.02 & 0.02 & 0 & 5 & 0 & 0 & 0.09 & 0.11 & 0 & 4 & 0 & 0 \\
\cline{1-20}
\multirow[t]{3}{*}{100} & 0.25 & 14.09 & 14.10 & 0 & 5 & 0 & 0 & 119.62 & 119.64 & 0 & 5 & 0 & 0 & 973.15 & 973.17 & 0 & 3 & 0 & 0 \\
 & 0.50 & 0.01 & 0.03 & 0 & 5 & 0 & 0 & 0.16 & 0.17 & 0 & 5 & 0 & 0 & 2.52 & 2.53 & 0 & 5 & 0 & 0 \\
 & 0.75 & 0.02 & 0.02 & 0 & 5 & 0 & 0 & 0.01 & 0.02 & 0 & 5 & 0 & 0 & 0.18 & 0.18 & 0 & 5 & 0 & 0 \\
\cline{1-20}
\multirow[t]{3}{*}{110} & 0.25 & 16.77 & 16.79 & 0 & 5 & 0 & 0 & 198.34 & 198.36 & 0 & 5 & 0 & 0 & 1106.35 & 1106.40 & 0 & 1 & 0 & 0 \\
 & 0.50 & 0.03 & 0.04 & 0 & 5 & 0 & 0 & 0.20 & 0.21 & 0 & 5 & 0 & 0 & 2.38 & 2.40 & 0 & 5 & 0 & 0 \\
 & 0.75 & 0.01 & 0.01 & 0 & 5 & 0 & 0 & 0.01 & 0.02 & 0 & 5 & 0 & 0 & 0.15 & 0.15 & 0 & 5 & 0 & 0 \\
\cline{1-20}
\multirow[t]{3}{*}{120} & 0.25 & 21.46 & 21.48 & 0 & 5 & 0 & 0 & 190.33 & 190.35 & 0 & 5 & 0 & 0 & 600.24 & 600.28 & 0 & 1 & 0 & 0 \\
 & 0.50 & 0.05 & 0.05 & 0 & 5 & 0 & 0 & 0.09 & 0.11 & 0 & 5 & 0 & 0 & 2.50 & 2.52 & 0 & 5 & 0 & 0 \\
 & 0.75 & 0.01 & 0.02 & 0 & 5 & 0 & 0 & 0.02 & 0.03 & 0 & 5 & 0 & 0 & 0.17 & 0.18 & 0 & 5 & 0 & 0 \\
\cline{1-20}
\multirow[t]{3}{*}{130} & 0.25 & 199.80 & 199.83 & 0 & 4 & 0 & 0 & 332.07 & 332.09 & 0 & 4 & 0 & 0 & 53.03 & 53.07 & 0 & 1 & 0 & 0 \\
 & 0.50 & 0.06 & 0.08 & 0 & 5 & 0 & 0 & 0.14 & 0.15 & 0 & 5 & 0 & 0 & 1.57 & 1.58 & 0 & 5 & 0 & 0 \\
 & 0.75 & 0.01 & 0.02 & 0 & 5 & 0 & 0 & 0.01 & 0.01 & 0 & 5 & 0 & 0 & 0.30 & 0.31 & 0 & 5 & 0 & 0 \\
\cline{1-20}
 & Mean & 6.14 & 6.21 & 0.36 & 4.26 & 0.36 & 0.0 & 20.97 & 25.55 & 1.0 & 3.69 & 0.29 & 0.0 & 76.59 & 144.10 & 2.03 & 1.85 & 0.03 & 0.36 \\
\cline{1-20}
\bottomrule
\end{tabular}
}
\end{table}


As observed in Table~\ref{experiment_sandwich_70_80_90}, feasible instances are heavily concentrated in settings with higher optional edge densities and lower graph orders. Naturally, as the optional edge density increases, the number of feasible instances also increases; this is expected, as a higher density expands the valid solution space (indeed, if the optional edge density is 1.0, the instance is trivially feasible). Conversely, the frequency of feasible instances diminishes as the graph order expands. Larger graphs inherently contain more odd holes and odd antiholes, thereby increasing the probability that at least one such structure cannot be resolved using the available optional edges. Finally, regarding the initial input graph, the number of feasible instances positively correlates with input density. This aligns with our observations from \textsc{MinPerfComp}, where denser initial graphs generally yielded more favorable performance."

Under the 0.90 optional edge density columns, several results are omitted, meaning fewer than five instances are reported for certain rows. In these specific cases, the operating system terminated the solver due to excessive memory consumption. Notably, these memory-intensive instances lie along a ``frontier region''---the structural boundary where predominantly feasible instances (the upper-right triangle of Table~\ref{experiment_sandwich_70_80_90}) transition into predominantly infeasible instances (the lower-left triangle). As is common in combinatorial feasibility problems, instances located near this transition boundary are notoriously the most computationally challenging to resolve, requiring significantly more memory and search effort to prove either feasibility or infeasibility.

\subsection{Performance of the Primal Heuristic (\texttt{IMH})} \label{sec:exp_imh}
\begin{sloppypar} 
In this section, we examine the performance of Algorithm \texttt{IterativeModification}-\texttt{Heuristic}. This algorithm is used as a primal heuristic to generate feasible solutions within the branch-and-cut algorithm, providing upper bounds and pruning the search space. However, both the branch-and-cut and the \texttt{IMH} can also be seen as standalone algorithms for generating perfect graphs. The \texttt{IMH} outputs a perfect graph while greedily minimizing the number of modifications. Although it typically yields suboptimal solutions with respect to the minimum editing objective, it is significantly more time efficient than the branch-and-cut algorithm for the purpose of obtaining a perfect graph. To specifically observe the runtime performance and the solution quality of the \texttt{IMH} relative to the branch-and-cut algorithm, we evaluated the heuristic on \textsc{MinPerfEdit}, with the same graphs used in the experiments presented in Section~\ref{sec:edit}.
\end{sloppypar}

Table~\ref{table:experiment_heuristic} reports the results of the \texttt{IMH} as a standalone algorithm for \erdosrenyi{}, $C_5$-free, and perturbed perfect graphs. For these instances, we compare the performance of the heuristic in terms of the minimum editing objective. To do so, we used the best lower bound at the end of the branch-and-cut algorithm achieved with the `H on C\&R' setting of the \texttt{IMH}. The `CP\_Gap(\%)' column represents the optimality gap of the branch-and-cut algorithm. The `Heur\_Gap(\%)' and `Time(s)' columns report the average objective gap and the average runtime of the \texttt{IMH}.

\begin{figure}[!htbp]
   \centering
   \includegraphics[width=0.9\textwidth]{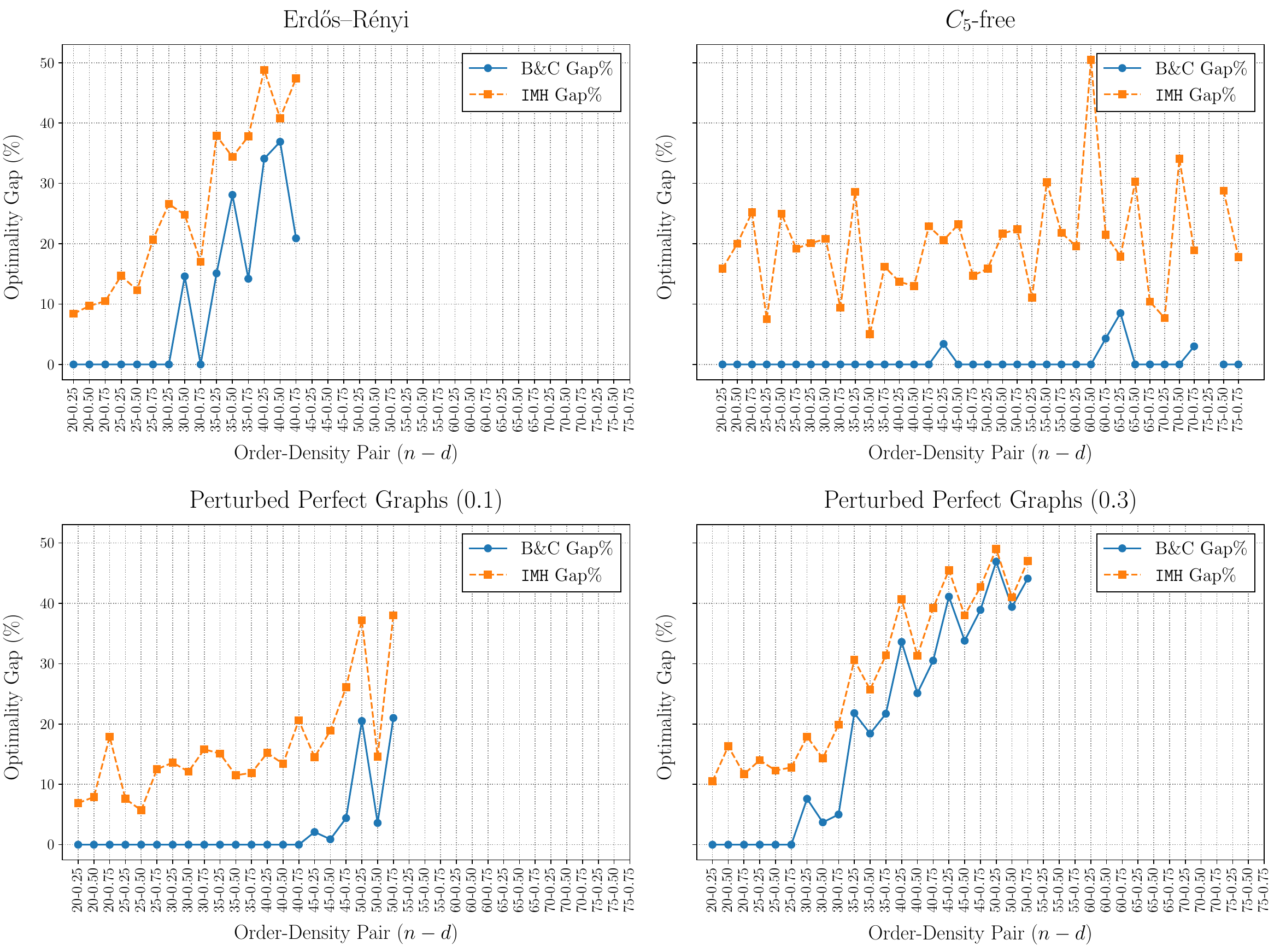}
   \caption{Optimality gap comparison between the branch-and-cut algorithm and the \texttt{IterativeModificationHeuristic} for \textsc{MinPerfEdit}.}
   \label{fig:imh_bc_gaps}
\end{figure}

Furthermore, Figure~\ref{fig:imh_bc_gaps} summarizes the optimality gaps of the \texttt{IMH} and the branch-and-cut algorithm across various graph types, based on the data in Table~\ref{table:experiment_heuristic}. We do not observe a significant difference in the optimality gaps, which suggests that the \texttt{IMH} outputs high-quality solutions for a heuristic method. The optimality gaps are especially close to those of the branch-and-cut algorithm for \erdosrenyi{} graphs and perturbed perfect graphs with a perturbation ratio of 0.3. Furthermore, the runtime of the \texttt{IMH} is substantially lower compared to that of the branch-and-cut algorithm, allowing it to be used efficiently for generating perfect graphs. However, the runtime of the \texttt{IMH} rises drastically as the graph order increases, which is due to the exponential growth in the number of odd holes and odd antiholes.

\begin{table}
\centering
\setlength{\tabcolsep}{4pt}
\caption{Experimental results of Algorithm~\texttt{IterativeModificationHeuristic} on several graph types graphs with varying orders and densities.}
\label{table:experiment_heuristic}
\resizebox{1\textwidth}{!}{
\begin{tabular}{cc|rrrr|rrrr|rrrr|rrrr}
\toprule
&  & \multicolumn{4}{c|}{\erdosrenyi{}} & \multicolumn{4}{c|}{$C_5$-free} & \multicolumn{4}{c|}{Perturbed Perfect Graphs (0.1)} & \multicolumn{4}{c}{Perturbed Perfect Graphs (0.3)} \\
\cmidrule(lr){3-6} \cmidrule(lr){7-10} \cmidrule(lr){11-14} \cmidrule(lr){15-18}
$n$ & $d$ & \makecell{CP\\Gap\%} & \makecell{\texttt{IMH}\\Gap\%} & \makecell{\texttt{IMH}\\Obj.} & \makecell{\texttt{IMH}\\Time(s)} & \makecell{CP\\Gap\%} & \makecell{\texttt{IMH}\\Gap\%} & \makecell{\texttt{IMH}\\Obj.} & \makecell{\texttt{IMH}\\Time(s)} & \makecell{CP\\Gap\%} & \makecell{\texttt{IMH}\\Gap\%} & \makecell{\texttt{IMH}\\Obj.} & \makecell{\texttt{IMH}\\Time(s)} & \makecell{CP\\Gap\%} & \makecell{\texttt{IMH}\\Gap\%} & \makecell{\texttt{IMH}\\Obj.} & \makecell{\texttt{IMH}\\Time(s)} \\
\midrule
\multirow[t]{3}{*}{20} & 0.25    & 0.0 & 8.4 & 8.8 & 0.0 & 0.0 & 15.9 & 4.2 & 0.0  & 0.0 & 6.9 & 5.8 & 0.0 & 0.0 & 10.5 & 14.4 & 0.0 \\
                        & 0.50   & 0.0 & 9.7 & 15.8 & 0.0 & 0.0 & 20.0 & 1.6 & 0.0  & 0.0 & 7.9 & 5.4 & 0.0 & 0.0 & 16.3 & 14.6 & 0.0 \\
                        & 0.75   & 0.0 & 10.5 & 9.4 & 0.0 & 0.0 & 25.2 & 3.8 & 0.0  & 0.0 & 17.9 & 6.8 & 0.0 & 0.0 & 11.7 & 14.8 & 0.0 \\
\cline{1-18}          
\multirow[t]{3}{*}{25} & 0.25    & 0.0 & 14.7 & 17.6 & 0.0 & 0.0 & 7.5 & 3.8 & 0.0  & 0.0 & 7.6 & 11.6 & 0.0 & 0.0 & 14.0 & 26.8 & 0.0 \\
                        & 0.50   & 0.0 & 12.3 & 30.8 & 0.0 & 0.0 & 25.0 & 4.0 & 0.0  & 0.0 & 5.7 & 8.8 & 0.0 & 0.0 & 12.3 & 26.2 & 0.0 \\
                        & 0.75   & 0.0 & 20.7 & 18.8 & 0.0 & 0.0 & 19.2 & 7.0 & 0.0  & 0.0 & 12.5 & 11.6 & 0.0 & 0.0 & 12.8 & 25.4 & 0.0 \\
\cline{1-18}
\multirow[t]{3}{*}{30} & 0.25    & 0.0 & 26.6 & 32.4 & 0.0 & 0.0 & 20.1 & 12.2 & 0.0  & 0.0 & 13.6 & 20.0 & 0.0 & 7.6 & 17.9 & 47.8 & 0.0 \\
                        & 0.50   & 14.6 & 24.8 & 56.6 & 0.0 & 0.0 & 20.8 & 5.8 & 0.0  & 0.0 & 12.1 & 19.4 & 0.0 & 3.7 & 14.3 & 46.0 & 0.0 \\
                        & 0.75   & 0.0 & 17.0 & 36.8 & 0.0 & 0.0 & 9.4 & 5.8 & 0.0  & 0.0 & 15.8 & 21.0 & 0.0 & 5.0 & 19.9 & 48.0 & 0.0 \\
\cline{1-18}
\multirow[t]{3}{*}{35} & 0.25    & 15.1 & 37.9 & 62.6 & 0.0 & 0.0 & 28.6 & 20.4 & 0.0  & 0.0 & 15.1 & 25.0 & 0.0 & 21.8 & 30.6 & 73.0 & 0.0 \\
                        & 0.50   & 28.1 & 34.4 & 86.8 & 0.0 & 0.0 & 5.0 & 1.8 & 0.0  & 0.0 & 11.5 & 27.2 & 0.0 & 18.4 & 25.7 & 68.8 & 0.0 \\
                        & 0.75   & 14.2 & 37.8 & 61.0 & 0.0 & 0.0 & 16.2 & 15.0 & 0.0  & 0.0 & 11.9 & 28.0 & 0.0 & 21.7 & 31.4 & 76.0 & 0.0 \\
\cline{1-18}
\multirow[t]{3}{*}{40} & 0.25    & 34.1 & 48.8 & 88.2 & 0.0 & 0.0 & 13.7 & 12.6 & 0.0  & 0.0 & 15.2 & 41.2 & 0.0 & 33.6 & 40.7 & 108.2 & 0.0 \\
                        & 0.50   & 36.9 & 40.8 & 123.2 & 0.0 & 0.0 & 13.0 & 3.4 & 0.0  & 0.0 & 13.4 & 36.4 & 0.0 & 25.1 & 31.3 & 93.6 & 0.0 \\
                        & 0.75   & 20.9 & 47.4 & 82.6 & 0.0 & 0.0 & 22.9 & 16.0 & 0.0  & 0.0 & 20.6 & 44.8 & 0.0 & 30.5 & 39.2 & 106.0 & 0.0 \\
\cline{1-18}
\multirow[t]{3}{*}{45} & 0.25    & --- & --- & 119.8 & 0.6 & 3.4 & 20.6 & 17.2 & 0.0  & 2.1 & 14.5 & 52.2 & 0.0 & 41.1 & 45.5 & 145.6 & 0.0 \\
                        & 0.50   & --- & --- & 164.2 & 0.0 & 0.0 & 23.2 & 8.2 & 0.0  & 0.9 & 18.9 & 53.0 & 0.0 & 33.8 & 38.0 & 126.6 & 0.0 \\
                        & 0.75   & --- & --- & 115.8 & 0.4 & 0.0 & 14.7 & 13.2 & 0.0  & 4.4 & 26.1 & 57.6 & 0.0 & 38.9 & 42.7 & 134.4 & 0.0 \\
\cline{1-18}
\multirow[t]{3}{*}{50} & 0.25    & --- & --- & 169.2 & 3.0 & 0.0 & 15.9 & 16.0 & 0.2  & 20.5 & 37.2 & 81.6 & 0.0 & 46.9 & 49.0 & 188.8 & 1.0 \\
                        & 0.50   & --- & --- & 204.8 & 1.0 & 0.0 & 21.7 & 6.2 & 0.0  & 3.6 & 14.6 & 64.0 & 0.0 & 39.4 & 41.0 & 161.8 & 0.0 \\
                        & 0.75   & --- & --- & 170.6 & 2.6 & 0.0 & 22.4 & 17.0 & 0.0  & 21.0 & 38.0 & 82.2 & 0.0 & 44.1 & 47.0 & 180.6 & 1.0 \\
\cline{1-18}
\multirow[t]{3}{*}{55} & 0.25    & --- & --- & 217.2 & 13.8 & 0.0 & 11.1 & 14.6 & 1.4  & --- & --- & 93.4 & 1.2 & --- & --- & 237.8 & 4.0 \\
                        & 0.50   & --- & --- & 262.8 & 4.2 & 0.0 & 30.2 & 11.8 & 0.0  & --- & --- & 80.8 & 0.0 & --- & --- & 207.4 & 2.0 \\
                        & 0.75   & --- & --- & 201.6 & 14.0 & 0.0 & 21.8 & 14.4 & 0.8  & --- & --- & 95.0 & 0.8 & --- & --- & 230.4 & 3.4 \\
\cline{1-18}
\multirow[t]{3}{*}{60} & 0.25    & --- & --- & 274.6 & 56.0 & 0.0 & 19.6 & 16.8 & 4.0  & --- & --- & 122.6 & 5.0 & --- & --- & 293.4 & 11.2 \\
                        & 0.50   & --- & --- & 338.6 & 12.2 & 0.0 & 50.5 & 22.4 & 0.2  & --- & --- & 104.6 & 0.2 & --- & --- & 260.4 & 6.2 \\
                        & 0.75   & --- & --- & 277.0 & 53.8 & 4.3 & 21.5 & 19.4 & 4.2  & --- & --- & 114.6 & 2.8 & --- & --- & 290.6 & 11.2 \\
\cline{1-18}
\multirow[t]{3}{*}{65} & 0.25    & --- & --- & 364.2 & 101.8 & 8.5 & 17.9 & 21.8 & 12.4  & --- & --- & 147.2 & 7.6 & --- & --- & 368.4 & 23.0 \\
                        & 0.50   & --- & --- & 385.6 & 18.2 & 0.0 & 30.3 & 14.8 & 0.0  & --- & --- & 123.8 & 1.0 & --- & --- & 294.0 & 8.0 \\
                        & 0.75   & --- & --- & 371.2 & 102.2 & 0.0 & 10.4 & 13.2 & 5.2  & --- & --- & 145.2 & 9.0 & --- & --- & 361.2 & 19.4 \\
\cline{1-18}
\multirow[t]{3}{*}{70} & 0.25    & --- & --- & 434.8 & 362.2 & 0.0 & 7.7 & 17.6 & 15.0  & --- & --- & 187.4 & 49.8 & --- & --- & 428.0 & 59.6 \\
                        & 0.50   & --- & --- & 469.0 & 49.2 & 0.0 & 34.1 & 14.6 & 0.2  & --- & --- & 152.2 & 3.6 & --- & --- & 355.2 & 24.2 \\
                        & 0.75   & --- & --- & 435.8 & 337.6 & 3.0 & 18.9 & 26.2 & 29.0  & --- & --- & 183.0 & 29.2 & --- & --- & 432.2 & 55.2 \\
\cline{1-18}
\multirow[t]{3}{*}{75} & 0.25    & --- & --- & 524.2 & 1018.2 & --- & --- & 32.0 & 62.6  & --- & --- & 201.2 & 140.6 & --- & --- & 498.6 & 123.6 \\
                        & 0.50   & --- & --- & 554.4 & 113.8 & 0.0 & 28.8 & 15.2 & 0.2  & --- & --- & 171.0 & 6.8 & --- & --- & 430.0 & 54.6 \\
                        & 0.75   & --- & --- & 514.8 & 879.2 & 0.0 & 17.8 & 16.0 & 24.6  & --- & --- & 201.0 & 97.8 & --- & --- & 539.2 & 125.6 \\
\cline{1-18}
\multirow[t]{3}{*}{80} & 0.25    & --- & --- & 609.4 & 1824.4 & --- & --- & 25.6 & 88.8  & --- & --- & 268.2 & 442.8 & --- & --- & 613.2 & 274.0 \\
                        & 0.50   & --- & --- & 637.2 & 195.4 & --- & --- & 17.2 & 1.0  & --- & --- & 218.8 & 22.4 & --- & --- & 487.4 & 115.4 \\
                        & 0.75   & --- & --- & 637.2 & 2246.0 & --- & --- & 23.0 & 68.4  & --- & --- & 255.6 & 316.0 & --- & --- & 620.6 & 232.0 \\
\cline{1-18}
\bottomrule
\end{tabular}
}
\end{table}

Table~\ref{table:experiment_heuristic_dimacs} presents the results of the \texttt{IMH} on the DIMACS instances alongside the best outcomes obtained using the branch-and-cut algorithm (referenced from Table~\ref{experiment_edit_dimacs}). The comparative performance of the \texttt{IMH} is striking, as it achieves high-quality solutions in a fraction of the runtime. Most notably, the \texttt{IMH} successfully identified the exact optimal solution for five of the 22 DIMACS instances (marked in bold), accomplishing this in strictly less time than the branch-and-cut method in every case.

\begin{table}
\centering
\setlength{\tabcolsep}{4pt}
\caption{Experimental results of Algorithm~\texttt{IterativeModificationHeuristic} on several DIMACS instances.}
\label{table:experiment_heuristic_dimacs}
\resizebox{0.72\textwidth}{!}{
\begin{tabular}{lrrrr|rr|rr|}
                    & \multicolumn{1}{l}{} & \multicolumn{1}{l}{} & \multicolumn{1}{l}{}   & \multicolumn{1}{l|}{}   & \multicolumn{2}{c|}{Branch-and-Cut Best Result} & \multicolumn{2}{c|}{\texttt{IMH}}\\ \hline
Instance            & $n$                  & $m$                  & \#$OH$ & \#$OAH$ & Objective                     & Time(s)        & Objective                              & Time(s)                              \\ \hline
anna                & 138                  & 493                  & 202                    & 0                       & 15                            & 0.8            & 19              & 0      \\
david               & 87                   & 406                  & 953                    & 0                       & 16                            & 1.0            & 28              & 0      \\
\textbf{fpsol2.i.2} & 451                  & 8691                 & 30016                  & 0                       & \textbf{28}                   & 43.0           & \textbf{28}     & 14     \\
\textbf{fpsol2.i.3} & 425                  & 8688                 & 30016                  & 0                       & \textbf{28}                   & 35.0           & \textbf{28}     & 13     \\
\textbf{huck}       & 74                   & 301                  & 3                      & 0                       & \textbf{2}                    & 0.1            & \textbf{2}      & 0      \\
\textbf{inithx.i.1} & 864                  & 18707                & 16                     & 0                       & \textbf{1}                    & 239.7          & \textbf{1}      & 105    \\
\textbf{jean}       & 80                   & 254                  & 67                     & 0                       & \textbf{6}                    & 0.1            & \textbf{6}      & 0      \\
miles1500           & 128                  & 5198                 & 72532                  & 1811                    & 24                            & 3.4            & 25              & 2      \\
miles250            & 128                  & 387                  & 922                    & 0                       & 13                            & 28.1           & 22              & 25     \\
mulsol.i.2          & 188                  & 3885                 & 14336                  & 0                       & 25                            & 5.4            & 26              & 1      \\
mulsol.i.3          & 184                  & 3916                 & 14336                  & 0                       & 25                            & 3.3            & 26              & 1      \\
mulsol.i.4          & 185                  & 3946                 & 14336                  & 0                       & 25                            & 3.6            & 26              & 1      \\
mulsol.i.5          & 186                  & 3973                 & 14336                  & 0                       & 25                            & 2.2            & 26              & 1      \\
myciel3             & 11                   & 20                   & 31                     & 0                       & 4                             & 0.0            & 6               & 0      \\
myciel4             & 23                   & 71                   & 646                    & 0                       & 16                            & 0.2            & 24              & 0      \\
myciel5             & 47                   & 236                  & 17277                  & 0                       & 56                            & 189.6          & 96              & 0      \\
queen5\_5           & 25                   & 160                  & 600                    & 136                     & 31.48 -- 39 (19.29\%)         & 903.0          & 45              & 0      \\
queen6\_6           & 36                   & 290                  & 10160                  & 456                     & 61.82 -- 91 (32.06\%)         & 900.3          & 128             & 0      \\
queen7\_7           & 49                   & 476                  & 120288                 & 1056                    & 98.4 -- 253 (61.11\%)         & 924.1          & 274             & 4      \\
queen8\_8           & 64                   & 728                  & 1685760                & 1960                    & 145.6 -- 728 (80.0\%)         & 912.3          & 576             & 133    \\
zeroin.i.2          & 211                  & 3541                 & 102640                 & 0                       & 70                            & 84.0           & 73              & 6      \\
zeroin.i.3          & 206                  & 3540                 & 102640                 & 0                       & 70                            & 80.1           & 73              & 6      \\ \hline
\end{tabular}
}
\end{table}

\section{Conclusion}\label{sec:conclusion}

Perfect graph modification problems sit at the intersection of structural graph theory and combinatorial optimization, yet despite their theoretical depth and practical relevance---spanning scheduling, frequency assignment, and beyond---no exact or heuristic algorithms had been proposed for them prior to this work. We close this gap with our contributions.

We propose the first exact algorithms for the minimum perfect editing problem, the minimum perfect completion problem, and the perfect sandwich problem. Our approach formulates each as an integer program grounded in the Strong Perfect Graph Theorem, and solves it via a branch-and-cut algorithm that generates odd hole and odd antihole constraints on demand.

Algorithm \texttt{IterativeModificationHeuristic} is the first heuristic designed for the minimum perfect editing and completion problems. Beyond its primary role of providing strong upper bounds to the branch-and-cut algorithm, the \texttt{IMH} doubles as a practical method for generating random perfect graphs---a capability that is notably absent from the literature and opens a promising direction for benchmarking and algorithm design on perfect graphs.

Across hundreds of instances spanning \erdosrenyi{} graphs, $C_5$-free graphs, perturbed perfect graphs, and established DIMACS benchmarks, our methods solve these to proven optimality in seconds to minutes. Moreover, the \texttt{IMH} alone matches the branch-and-cut optimum on several DIMACS instances while running faster on higher graph orders. Furthermore, because the perfect sandwich problem is fundamentally a decision problem, it scales to significantly higher graph orders than its exact optimization counterparts.

\begin{sloppypar}
To support reproducibility and future research, all benchmark instances used in our experiments are publicly available \citep{erdem_github_modification}, along with the graph generation codes \citep{erdem_github_generation}. Moreover, we adapted Algorithm \texttt{FindOddHoles}, which enumerates odd holes and odd antiholes efficiently in practice, into a standalone version called \texttt{is\_perfect}. This ready-to-use perfect graph recognizer---featuring documented average runtimes across varying graph types, orders, and densities---is released as an independent open-source tool \citep{erdem_is_perfect}, serving as a valuable subroutine for any algorithm requiring odd hole detection or perfect graph recognition.
\end{sloppypar}

Looking ahead, the branch-and-cut paradigm introduced here extends naturally to other graph modification targets whenever the desired property admits a forbidden-subgraph characterization whose certificates can be found explicitly and encoded as linear constraints. We also identify the generation of random perfect graphs as a rich open problem. The \texttt{IMH} provides a first practical handle, but dedicated generators with theoretical guarantees remain an important direction for future work.


\textbf{Acknowledgments:}
This study is supported by the Scientific and Technological Research Council of Turkey (TUBITAK), Grant No. 122M452. We would like to thank Professor Ümit Işlak for his help regarding the calculation of the expected number of odd holes and odd antiholes in random \erdosrenyi{} graphs.

\bibliographystyle{informs2014}
\bibliography{references}

\end{document}